\documentclass[twocolumn,tighten]{aastex62}

\usepackage{amsmath}
\usepackage{amssymb}
\usepackage{xcolor}
\usepackage{hyperref}
\usepackage[caption=false]{subfig}
\usepackage{longtable}
\usepackage{afterpage}
\usepackage{xspace}
\usepackage{upgreek}
\usepackage{multirow}
\usepackage{cmap}

\usepackage{float}
\usepackage{footnote}

\graphicspath{{./}{figures/}}

\received{\today}
\revised{\today}
\accepted{XXX}
\submitjournal{ApJ}

\shorttitle{Breakthrough Listen Observations of \tess\ Stars}
\shortauthors{R. Traas et. al.}

\begin{document}

\newcommand{\tseti}{{{turbo}SETI}}
\newcommand{\tiddalik}{\textsc{Tiddalik} }
\newcommand{\blimpy}{\textsc{Blimpy} }
\newcommand{\Hzs}{Hz\,s$^{-1}$}
\newcommand{\EventBW}{$\Delta\nu_{\rm{event}}$}
\newcommand{\tess}{\emph{TESS}}
\newcommand{\dock}{\textsc{Docker }}
\newcommand{\tmux}{\emph{tmux }}
\newcommand{\snr}{$\texttt{S/N}$}

\newcommand{\ntoi}{{\color{black}28}}
\newcommand{\nexo}{{\color{black}66}}
\newcommand{\ntoitotal}{{\color{black}2120}}

\newcommand{\tsysLall}{{\color{black}18.46}}
\newcommand{\tsysL}{{\color{black}15.60}}
\newcommand{\tsysSall}{{\color{black}16.94}}
\newcommand{\tsysS}{{\color{black}14.80}}
\newcommand{\tsysCall}{{\color{black}22.42}}
\newcommand{\tsysC}{{\color{black}21.50}}
\newcommand{\tsysXall}{{\color{black}34.26}}
\newcommand{\tsysX}{{\color{black}30.80}}
\newcommand{\fminL}{{\color{black}5.304}}
\newcommand{\fminLall}{{\color{black}6.275}}
\newcommand{\fminS}{{\color{black}5.032}}
\newcommand{\fminSall}{{\color{black}5.759}}
\newcommand{\fminC}{{\color{black}7.310}}
\newcommand{\fminCall}{{\color{black}7.623}}
\newcommand{\fminX}{{\color{black}10.470}}
\newcommand{\fminXall}{{\color{black}11.648}}
\newcommand{\eirpminL}{{\color{black}491.6}}
\newcommand{\eirpminLall}{{\color{black}581.6}}
\newcommand{\eirpminS}{{\color{black}466.4}}
\newcommand{\eirpminSall}{{\color{black}533.7}}
\newcommand{\eirpminC}{{\color{black}677.5}}
\newcommand{\eirpminCall}{{\color{black}706.5}}
\newcommand{\eirpminX}{{\color{black}970.5}}
\newcommand{\eirpminXall}{{\color{black}1080.0}}

\newcommand{\eirplimit}{{\color{black}$4.9\times10^{14}$\,W}}

\newcommand{\sefdL}{{\color{black}7.7251}}
\newcommand{\sefdLall}{{\color{black}9.1393}}
\newcommand{\sefdS}{{\color{black}7.32896}}
\newcommand{\sefdSall}{{\color{black}8.3876}}
\newcommand{\sefdC}{{\color{black}10.6468}}
\newcommand{\sefdCall}{{\color{black}11.1032}}
\newcommand{\sefdX}{{\color{black}15.2522}}
\newcommand{\sefdXall}{{\color{black}16.9653}}

\newcommand{\nstar}{{\color{black}1327 }}
\newcommand{\nstaradd}{{\color{black}641 }}  

\newcommand{\nobsgbtS}{{\color{black}6456 }}
\newcommand{\nobsgbtL}{{\color{black}6042 }}
\newcommand{\nobsgbtSL}{{\color{black}12504 }}

\newcommand{\nstargbt}{{\color{black}1138 }}
\newcommand{\nstargbtSL}{{\color{black}749 }}
\newcommand{\nstargbtL}{{\color{black}107}}
\newcommand{\nstargbtS}{{\color{black}107}}
\newcommand{\nstargbtC}{{\color{black}110}}
\newcommand{\nstargbtX}{{\color{black}105}}
\newcommand{\nstargbtsum}{{\color{black}429}}

\newcommand{\ncadencegbt}{{\color{black}2089 }}
\newcommand{\ncadencegbtL}{{\color{black}28}}
\newcommand{\ncadencegbtS}{{\color{black}28}}
\newcommand{\ncadencegbtC}{{\color{black}28}}
\newcommand{\ncadencegbtX}{{\color{black}29}}
\newcommand{\ncadencegbtsum}{{\color{black}113}}
\newcommand{\nfiles}{{\color{black}696}}

\newcommand{\neventsum}{{\color{black}5361}}
\newcommand{\neventL}{{\color{black}616}}
\newcommand{\neventS}{{\color{black}2842}}
\newcommand{\neventX}{{\color{black}1634}}
\newcommand{\neventC}{{\color{black}269}}

\newcommand{\ncand}{{\color{black}651}}
\newcommand{\ncandL}{{\color{black}630}}
\newcommand{\ncandS}{{\color{black}15}}
\newcommand{\ncandC}{{\color{black}6}}
\newcommand{\ncandx}{{\color{black}0}}

\newcommand{\nhrpks}{483.0}
\newcommand{\nhrgbtL}{14.5}
\newcommand{\nhrgbtS}{14.5}
\newcommand{\nhrgbtC}{14.5}
\newcommand{\nhrgbtX}{14.5}
\newcommand{\nhrgbtsum}{58.0}

\newcommand{\nobspks}{{\color{black}5796}\xspace}
\newcommand{\nstarpks}{{\color{black}189}\xspace}
\newcommand{\nstarpksIsaacson}{{\color{black}183}\xspace}
\newcommand{\nstarpksExtra}{{\color{black}6}\xspace}
\newcommand{\ncadencepksdb}{{\color{black}990}\xspace}
\newcommand{\ncadencepks}{{\color{black}966}\xspace}

\newcommand{\nevent}{{\color{black} zero candidates}\xspace}
\newcommand{\neventpks}{{\color{black}77\xspace}}
\newcommand{\neventgbtL}{{\color{black}15998}\xspace}
\newcommand{\neventgbtS}{{\color{black}5102}\xspace}

\newcommand{\neventpksnstar}{20\xspace}  
\newcommand{\neventgbtLnstar}{831\xspace}
\newcommand{\neventgbtSnstar}{511\xspace}

\newcommand{\ngroupspks}{{\color{black}60}\xspace}
\newcommand{\ngroupsgbtL}{{\color{black}4522}\xspace}
\newcommand{\ngroupsgbtS}{{\color{black}1572}\xspace}

\newcommand{\nfinalpks}{0\xspace}
\newcommand{\nfinalL}{0\xspace}
\newcommand{\nfinalS}{0\xspace}

\newcommand{\nhitspks}{4.45M\xspace}
\newcommand{\nhitsgbtL}{{\color{black}37.14M}\xspace}
\newcommand{\nhitsgbtS}{{\color{black}10.12M}\xspace}

\newcommand{\nhitsL}{{\color{black}189321}}
\newcommand{\nhitsS}{{\color{black}20673}}
\newcommand{\nhitsC}{{\color{black}1054}}
\newcommand{\nhitsX}{{\color{black}38484}}
\newcommand{\nhitssum}{{\color{black}249532}}

\newcommand{\neventsL}{{\color{black}34727}}
\newcommand{\neventsS}{{\color{black}5485}}
\newcommand{\neventsC}{{\color{black}269}}
\newcommand{\neventsX}{{\color{black}2794}}
\newcommand{\neventssum}{{\color{black}43275}}

\newcommand{\nhitsgbtLzerodrift}{21.90M\xspace}
\newcommand{\nhitsgbtLnegativedrift}{13.9M\xspace}
\newcommand{\nhitsgbtLpositivedrift}{1.37M\xspace}

\newcommand{\nhitsgbtSzerodrift}{7.36M\xspace}
\newcommand{\nhitsgbtSpositivedrift}{1.21M\xspace}
\newcommand{\nhitsgbtSnegativedrift}{1.55M\xspace}

\newcommand{\tfmL}{{\color{black}522}}
\newcommand{\tfmS}{{\color{black}533}}
\newcommand{\tfmC}{{\color{black}478}}
\newcommand{\tfmX}{{\color{black}1089}}

\newcommand{\tfmLcad}{{\color{black}1995}}
\newcommand{\tfmScad}{{\color{black}2038}}
\newcommand{\tfmCcad}{{\color{black}1878}}
\newcommand{\tfmXcad}{{\color{black}4675}}

\newcommand{\tfmCombined}{{\color{black}1072}}

\newcommand{\ptransL}{{\color{black}$12.72\%$}}
\newcommand{\ptransS}{{\color{black}$12.72\%$}}
\newcommand{\ptransC}{{\color{black}$12.72\%$}}
\newcommand{\ptransX}{{\color{black}$12.72\%$}}
\newcommand{\ptrans}{{\color{black}$12.72\%$}}

\newcommand{\nevgL}{{\color{black}135}}
\newcommand{\nevgS}{{\color{black}2}}
\newcommand{\nevgC}{{\color{black}1}}
\newcommand{\nevgX}{{\color{black}0}}
\newcommand{\nssL}{{\color{black}20}}
\newcommand{\nssS}{{\color{black}0}}
\newcommand{\nssC}{{\color{black}0}}
\newcommand{\nssX}{{\color{black}0}}

\newcommand{\dataVol}{{\color{black} 24 TB}}
\newcommand{\totalSkyTime}{{\color{black} 3390 min}}


\newcommand{\BI}{\textit{Breakthrough Initiatives} }
\newcommand{\BLI}{\textit{Breakthrough Listen Initiative} }
\newcommand{\bl}{\textit{Breakthrough Listen} }
\newcommand{\BL}{BL\xspace}

\title{The Breakthrough Listen Search for Intelligent Life: \\ 
 Searching for Technosignatures in Observations of \tess\ Targets of Interest} 


\newcommand{\UWL}{Department of Physics, University of Wisconsin - La Crosse, 1725 State Street, La Crosse, WI 54601, USA}
\newcommand{\UCB}{Department of Astronomy,  University of California Berkeley, Berkeley CA 94720}
\newcommand{\SSL}{Space Sciences Laboratory, University of California, Berkeley, Berkeley CA 94720}
\newcommand{\SWIN}{Centre for Astrophysics \& Supercomputing, Swinburne University of Technology, Hawthorn, VIC 3122, Australia}
\newcommand{\GBT}{Green Bank Observatory,  West Virginia, 24944, USA}
\newcommand{\OXF}{Astronomy Department, University of Oxford, Keble Rd, Oxford, OX13RH, United Kingdom}
\newcommand{\NIJ}{Department of Astrophysics/IMAPP,Radboud University, Nijmegen, Netherlands}
\newcommand{\ATNF}{Australia Telescope National Facility, CSIRO, PO Box 76, Epping, NSW 1710, Australia}
\newcommand{\HOU}{Hellenic Open University, School of Science \& Technology, Parodos Aristotelous, Perivola Patron, Greece}
\newcommand{\Hillsdale}{Hillsdale College, 33 E College St, Hillsdale, MI 49242}

 \newcommand{\USQ}{University of Southern Queensland, Toowoomba, QLD 4350, Australia}

\newcommand{\SETI}{SETI Institute, Mountain View, California}
\newcommand{\KZA}{University of Malta, Institute of Space Sciences and Astronomy}
\newcommand{\PWJD}{The Breakthrough Initiatives, NASA Research Park, Bld. 18, Moffett Field, CA, 94035, USA}
\newcommand{\COL}{Department of Astronomy, Columbia University, 550 West 120th Street, New York, NY 10027, USA}
\newcommand{\PENN}{Department of Astronomy and Astrophysics, Pennsylvania State University, University Park PA 16802}

\newcommand{\mgo}{M\&P }
\newcommand{\refsec}[1]{Section~\ref{#1}}
\newcommand{\reffig}[1]{Figure~\ref{#1}}
\newcommand{\reftab}[1]{Table~\ref{#1}}

\newcommand{\ee}[1]{\textbf{\color{blue} EE: #1}} 
\newcommand{\GSF}[1]{\textbf{\color{red} GSF: #1}} 
\newcommand{\dc}[1]{\textbf{\color{orange} DC: #1}} 
\newcommand{\dcp}[1]{\textbf{\color{purple} DCP: #1}} 
\newcommand{\todo}[1]{\textbf{\color{purple} TODO: #1}} 
\newcommand{\bcl}[1]{\textbf{\color[rgb]{0.2,0,0.5} BCL: #1}} 
\newcommand{\vg}[1]{\textbf{\color{green} VG: #1}} 
\newcommand{\ngi}[1]{\textbf{\color{cyan} NG: #1}} 
\newcommand{\diff}[1]{\textbf{\color{purple} #1}} 

\newcommand{\apvs}[1]{\textbf{\color{green} APVS: #1}} 

\correspondingauthor{Raffy Traas}
\email{traas5694@uwlax.edu\\ raffytraas14@gmail.com}

\author{Raffy Traas}
\affiliation{\UWL}
\affiliation{\UCB}

\author[0000-0003-4823-129X]{Steve Croft}
\affiliation{\UCB}
\affiliation{\SETI}





\author[0000-0002-8604-106X]{Vishal Gajjar}
\affiliation{\UCB}

\author[0000-0002-0531-1073]{Howard Isaacson}
\affiliation{\UCB}
\affiliation{\USQ}


\author{Matt Lebofsky}
\affiliation{\UCB}

\author{David H.\ E.\ MacMahon}
\affiliation{\UCB}


\author{Karen Perez}
\affiliation{\COL}

\author[0000-0003-2783-1608]{Danny C.\ Price}
\affiliation{\UCB}
\affiliation{\SWIN}

\author{Sofia Sheikh}
\affiliation{\PENN}
\affiliation{\UCB}

\author[0000-0003-2828-7720]{Andrew P.\ V.\ Siemion}
\affiliation{\UCB}
\affiliation{\SETI}
\affiliation{\NIJ}
\affiliation{\KZA}

\author{Shane Smith}
\affiliation{\Hillsdale}


\author{Jamie Drew}
\affiliation{\PWJD}

\author{S. Pete Worden}
\affiliation{\PWJD}

\begin{abstract}
Exoplanetary systems are prime targets for the Search for Extraterrestrial Intelligence (SETI).  With the recent uptick in the identification of candidate and confirmed exoplanets through the work of missions like the Transiting Exoplanet Survey Satellite (\tess), we are beginning to understand that Earth-like planets are common.  In this work, we extend the Breakthrough Listen (BL) search for extraterrestrial intelligence to include targeted searches of stars identified by \tess\ as potential exoplanet hosts. We report on $\ncadencegbtsum$ 30-min cadence observations collected for 28 targets selected from the \tess\ Input Catalog (TIC) from among those identified as containing signatures of transiting planets. The targets were searched for narrowband signals from 1 -- 11\,GHz using the \tseti\ \citep{Enriquez:2017, tseti_paper} pipeline architecture modified for compatibility with the Google Cloud environment.  Data were searched for drift rates of $\displaystyle \pm4$\,Hz\,/\,s above a minimum signal-to-noise threshold of $10$, following the parameters of previous searches conducted by \citet{Price:2020} and \citet{Enriquez:2017}.  The observations presented in this work establish some of the deepest limits to date over such a wide band (1 -- 11\,GHz) for life beyond Earth.  We determine that fewer than \ptransL\ of the observed targets possess transmitters operating at these frequencies with an Equivalent Isotropic Radiated Power (EIRP) greater than our derived threshold of \eirplimit.
\end{abstract}



\section{Introduction}\label{sec:introduction}
\subsection{SETI Methods}\label{subsec:seti-methods}
\begin{table*}[t]
\begin{center}
\caption{\label{tab:receiver-specifications} Observation parameters}
\begin{tabular}{cccccc}
\hline
Receiver & Frequency & $T_{\rm{sys}}$ & SEFD & $10\sigma$ Minimum Flux Density& $10\sigma$ Minimum EIRP\\
         &   (GHz)   &    (K)         & (Jy) & (Jy) & ($10^{12} W$)\\
\hline
\hline
L-band & 1.10--1.90  &  \tsysL        & \sefdL   &  \fminL                        &     \eirpminL\\
S-band & 1.80--2.80  &  \tsysS        & \sefdS   &  \fminS                        &     \eirpminS\\
C-band & 4.00--7.80  &  \tsysC        & \sefdC   &  \fminC                        &     \eirpminC\\
X-band & 7.80--11.20 &  \tsysX        & \sefdX    &  \fminX                        &     \eirpminX\\
\hline
\end{tabular}
\end{center}
\end{table*}
The Search for Extraterrestrial Intelligence (SETI) is the endeavor to discover the existence of intelligent life elsewhere in the universe -- one of the longest standing unanswered questions in science.  There are three traditional techniques to detecting intelligent life.  The first method is \emph{in situ} sampling, where studies are conducted at the site of interest.  While this approach would provide us with results containing the least amount of uncertainty, it is infeasible in practice when studying extrasolar objects due to the vast interstellar distances involved.  This brings about the need for techniques that can indirectly study sites of interest, which motivates the remaining two strategies: remote \emph{biosignature} and \emph{technosignature} searches.  Technosignature searches are concerned with the detection of evidence of a technologically sophisticated civilization through their city lights, atmospheric pollution, satellites, powerful transmitters, etc.  Traditionally, the majority of technosignature studies search for radio wave transmissions \citep{Cocconi1959} typically in the $1-10$\,GHz frequency range.  Biosignature searches use spectroscopy to detect planets with atmospheres or surfaces suitable for hosting life.  This method is capable of detecting the presence of organic compounds in planetary atmospheres and surfaces, thus indicating the possibility of life; however, it is challenging to unequivocally determine if those compounds derive from the emissions of living organisms or byproducts of inorganic processes.  More importantly, in the case that a biosignature search indisputably detects the presence of life, it remains uncertain as to whether these organisms are complex, or even technological.  In this regard, technosignature searches can be seen as a more holistic approach than biosignatures because, in principle, the detection of a technosignature would imply the existence of life.  Moreover, whereas even the next generation of telescopes  will only be able to probe nearby stars for biosignatures, searching for technosignatures allows the search to extend out to a much larger number of stars.

\begin{figure*}[ht]
\begin{center}
\includegraphics[width=2\columnwidth]{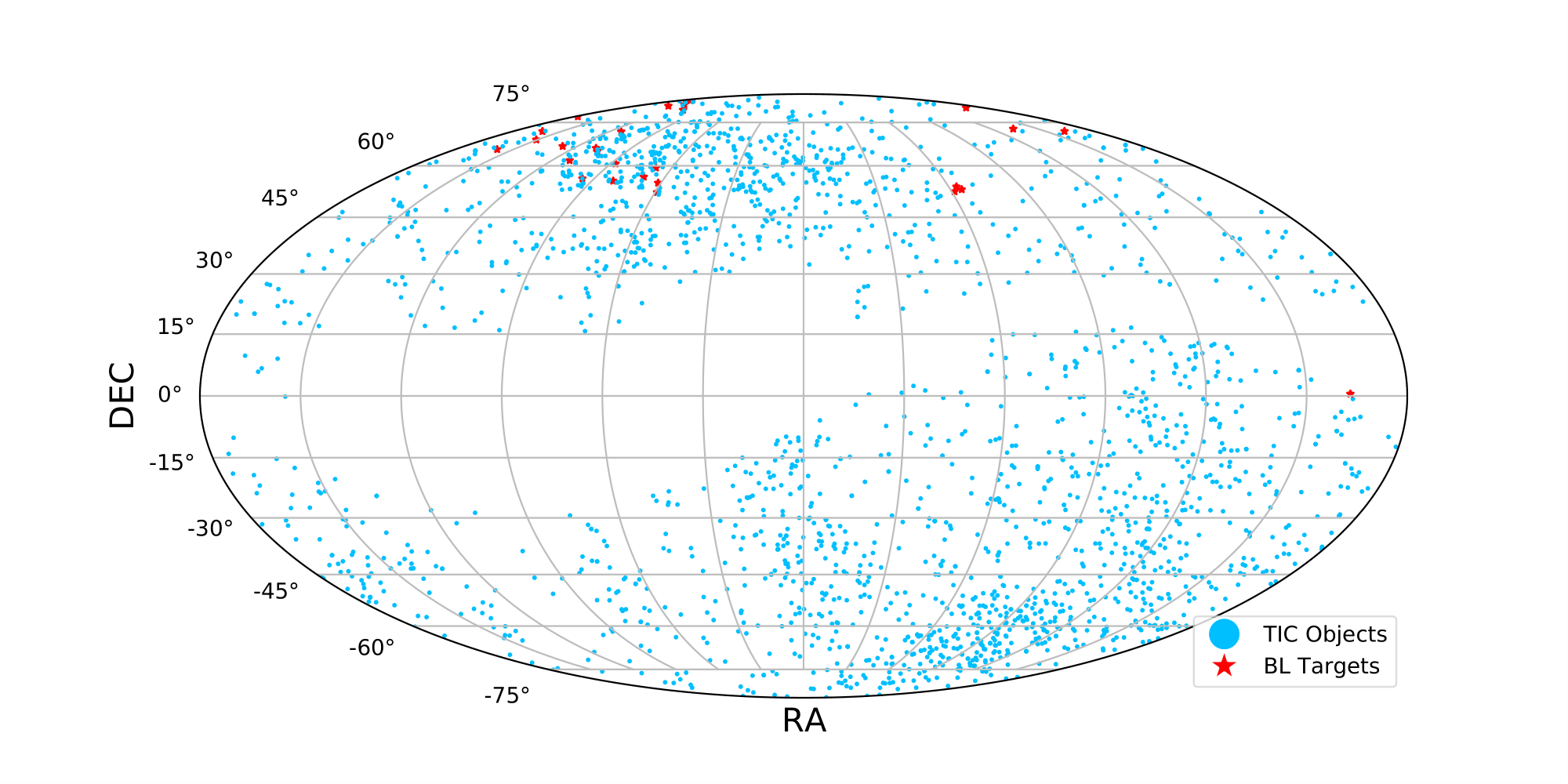}
\protect\caption{Distribution of BL targets analyzed in this paper, and a subset of existing TIC sources with associated TOI objects in equatorial coordinates.  The subset was collected from resources provided by the Exoplanet Follow-up Observing Program for \tess\ (ExoFOP-TESS).  The \tess\ project uses an all-sky survey approach to find transiting exoplanets.
\label{fig:allsky-tic}
}
\end{center}
\end{figure*}

\subsection{Breakthrough Listen}\label{subsec:BL}
Breakthrough Listen \citep{Worden:2017} is a ten-year initiative purposed towards the detection of technosignatures.  Launched in 2015, it is equipped with the most capable tools and resources with which to conduct a SETI search, and constitutes the most comprehensive search to date \citep{Enriquez:2017,Price:2020}.  BL conducts searches at optical wavelengths as well, by using facilities such as the Automated Planet Finder (APF) to search for laser emission lines present in stellar spectra \citep{Lipman2019}, and the VERITAS telescope to look for nanosecond optical pulses.

\subsection{Exoplanets}\label{subsec:exoplanets}
Exoplanets are promising locations to search for technosignatures \citep[e.g.,][]{2018ApJ...856...31T,Tremblay:2020}. Earth harbors intelligent life and is itself an exoplanet from an extrasolar perspective, and this naturally leads us to presume the possibility of life having emerged in a similar environment. Besides an exoplanet residing within the habitable zone of its host star, we are more likely to assume it might harbor intelligent life if its properties are similar to that of Earth's.  In the case that intelligent life eventually develops on these other worlds and become at least as technologically advanced as ourselves, they might possess some type of radio wave transmitting system, which forms the basis for many of our own telecommunications networks. Should such civilizations exist, we should be able to detect their presence as their signals inevitably spill out into space. While Breakthrough Listen has performed searches of specific exoplanet systems \citep[e.g.,][]{perez2020breakthrough, sofia_earth_transit_zone}, observing a larger sample can provide a limit on the number of nearby exoplanets on which radio transmitters aim signals directly toward Earth or output a sufficient amount of isotropic power to be detectable.

With the current surge in exoplanet discovery work being made by the \emph{NASA Transiting Exoplanet Survey Satellite} (\tess), the number of confirmed exoplanets has increased dramatically alongside the number of stars identified as having signatures of transiting planet candidates, making a technosignature search of these objects particularly timely.  Objects in the \tess\ Input Catalog (TIC), assembled from several existing catalogs as detailed by \citet{Stassun_2019}, are observed by \tess, and those identified as having signatures of transiting planets are subsequently registered into the \tess\ Objects of Interest (TOI) catalog -- a list of TIC targets marked for follow-up observation.  The recent completion of the \tess\ primary mission has found $\nexo$ confirmed exoplanets and $\ntoitotal$ targets of interest\footnote{\url{https://tess.mit.edu/publications/}}.

A collaboration between Breakthrough Listen and scientists working on the \tess\ mission has initiated the search for extraterrestrial intelligence in \tess\ target candidates.
Removing those listed as false positives, Breakthrough Listen observes objects present in the TOI catalog using the Exoplanet Follow-up Observing Program for \tess\ (ExoFOP-\tess\footnote{\url{https://exofop.ipac.caltech.edu/tess/view_toi.php}}).  The Robert C. Byrd Green Bank Observatory (hereafter GBT), a BL facility, observes targets above a declination of $-20\degr$, resulting in an observation queue of 964 objects.  
The \ntoi\ targets included in this work are those that, to date, have been observed at all four of L-, S-, C-, and X-bands with the GBT, and are presented in \reffig{fig:allsky-tic}.  This work searches more targets across $1 - 11$\,GHz than any previous BL search \citep[e.g.,][]{oumuamua_paper}.  A summary of the targets is given in Appendix~\ref{sec:targets}, from which the information included is sourced from the ExoFOP-\tess.

Breakthrough Listen maintains a non-prescriptive approach to the targets it observes \citep[see, e.g.,][]{Isaacson:2017, lacki:2020, Gajjar:2021}.  Although planetary systems are common, this sample adds significant value by extending the search to planetary systems with known transiting exoplanets.

Planetary systems are common, with the most common stars in the universe estimated to host several planets each \citep{hsu2020occurrence}. Given this fact, we should be clear that these TOI systems are not interesting solely because they host planets; they are interesting because of the unique relative geometry that a transiting planet implies \citep{sheikh2020nine}. If \tess\ discovers a transiting planet, then Earth is necessarily in the ecliptic of that planetary system. This geometry leads to a higher likelihood of observing both intentional (beacon) technosignatures and unintentional (leakage) technosignatures from that system. For example, an extraterrestrial intelligence (ETI) may preferentially send beacons along its ecliptic, knowing that its presence is most easily detected in that region of their sky due to their transit. Alternatively, the ETI may have similar interplanetary assets to those on Earth (e.g., radar to track asteroids in their system, spacecraft on or around other bodies in their planetary system), which would likely be distributed along the ecliptic; strong radio leakage may be produced preferentially in the ecliptic due to these activities. Finally, an ETI that has settled multiple bodies in the same system would likely have strong and frequent communications between them. Again, we would expect these communications to be roughly aligned with the ecliptic, increasing the possibility that we pick up spillover radio transmissions on Earth.

\section{Observations}\label{sec:observations}
Observations were conducted using the 100-m dish of the GBT and were recorded in an ABACAD cadence following the previous searches of \citet{Enriquez:2017} and \citet{Price:2020} in which each observation of a primary (`ON') target is alternated with observations of secondary (`OFF') targets. Each observation has a 5\,min duration, amounting to 30\,min for a full cadence of 6 observations, for a total of over \totalSkyTime\ on the sky. Data are recorded in accordance with the standard Breakthrough Listen strategy detailed by \citet{Lebofsky:2019}.  Details of the receivers are listed in \reftab{tab:receiver-specifications}.

\section{Analysis}\label{sec:analysis}
\subsection{De-Doppler Pipeline}\label{sec:dedoppler-pipeline}

The fine-frequency resolution data products, with a channelization of 2.7\,Hz \citep{Lebofsky:2019}, amounting to a total of \dataVol, were analyzed for the presence of narrowband drifting signals using the \tseti\footnote{\url{https://github.com/UCBerkeleySETI/turbo_seti}} \citep{Enriquez:2017, tseti_paper} pipeline.  The datasets analyzed are available from the Breakthrough Listen Open Data Archive\footnote{\url{http://seti.berkeley.edu/opendata}}.

TurboSETI is a Doppler drift search algorithm designed to search for narrowband drifting signals. Detecting narrowband signals in regions crowded with radio frequency interference (RFI), and over wide ranges of Doppler drift rates, is challenging \citep[see, e.g.,][]{Margot:2021}. Detection efficiencies for certain kinds of signals may be lower than expected, including for signals that are more complex than the simple assumption of a narrowband tone, and for signals with high drift rates. Minimum flux densities presented in \reftab{tab:receiver-specifications} are for a simple low-drift-rate tone. Work is ongoing to enhance the capabilities of \tseti, including better handling of signals spread across multiple frequency channels, and quantifying detection efficiency using signal injection and recovery\footnote{\url{https://github.com/krishnabhattaram/TurboSETIRetrieval}}. In addition, we are exploring approaches, including machine learning \citep{2020PASP..132k4501B}, that offer improved performance in regions crowded with RFI. Nevertheless, \tseti\ successfully finds the Voyager 1 spacecraft (an ``extraterrestrial'' transmitter that is a good stand-in for a real technosignature) in a blind search\footnote{\url{https://github.com/elanlavie/VoyagerTutorialRepository}}, and can be used to place useful limits on the prevalence of technosignatures in the dataset presented here.

Prior to this work, the Breakthrough Listen Doppler search pipeline was executed on the compute nodes at the UC Berkeley Data Center, where data to be processed was stored on disk.  This work is the result of an endeavor to migrate the \tseti\ pipeline onto the Google Cloud Platform (GCP).  Data were stored in a Google Cloud Storage (GCS) Bucket which was mounted to the file storage systems of each GCP Compute Engine to perform the analysis.  Each of the 20 instances was installed with a containerized version of \tseti\ as a \emph{Docker} \citep{docker} image\footnote{The \textsc{Docker} image used in this work can be found here: \url{https://hub.docker.com/r/rtraas/turbo-cloud}}.  This process is shown in \reffig{fig:turbo-cloud-architecture}.  In this manner, \tseti\ ran on a total of $\ncadencegbtsum$ full cadences, finding $\nhitssum$ hits over a total of $\nfiles$ individual observations.  Only complete, six-target cadence sets were analyzed.

\begin{table}[!h]
\captionsetup{size=footnotesize}
\caption{Summary Statistics} \label{tab:summary-stats}
\setlength\tabcolsep{0pt} 
\smallskip 
\begin{tabular*}{\columnwidth}{@{\extracolsep{\fill}}cccccr}
\hline
&   L-Band    &   S-Band    &   C-Band  &   X-Band  &   Overall\\
&   ($\%$)    & ($\%$) & ($\%$) & ($\%$) & ($\%$)\\
\hline
\hline
Hits  & 76 & 8        &   1       &   15      &     100\\
Events & 80 & 13 & 1 & 6 & 100\\
Candidates  & 97 & 2 & 1 & 0 & 100\\
Hit $\xrightarrow{}$ Event & 18 & 27 & 28 & 7 & 17\\
Event $\xrightarrow{}$ Candidate  & 2 & 0.27 & 2 & 0 & 2\\
Hit $\xrightarrow{}$ Candidate & 0.33 & 0.07 & 0.57 & 0 & 0.26\\
\hline
\end{tabular*}
\end{table}
\begin{figure*}
\centering 
\begin{tabular}{ccc}
\subfloat[Drift rate hit density, L-band]{%
  \includegraphics[width=0.7\columnwidth]{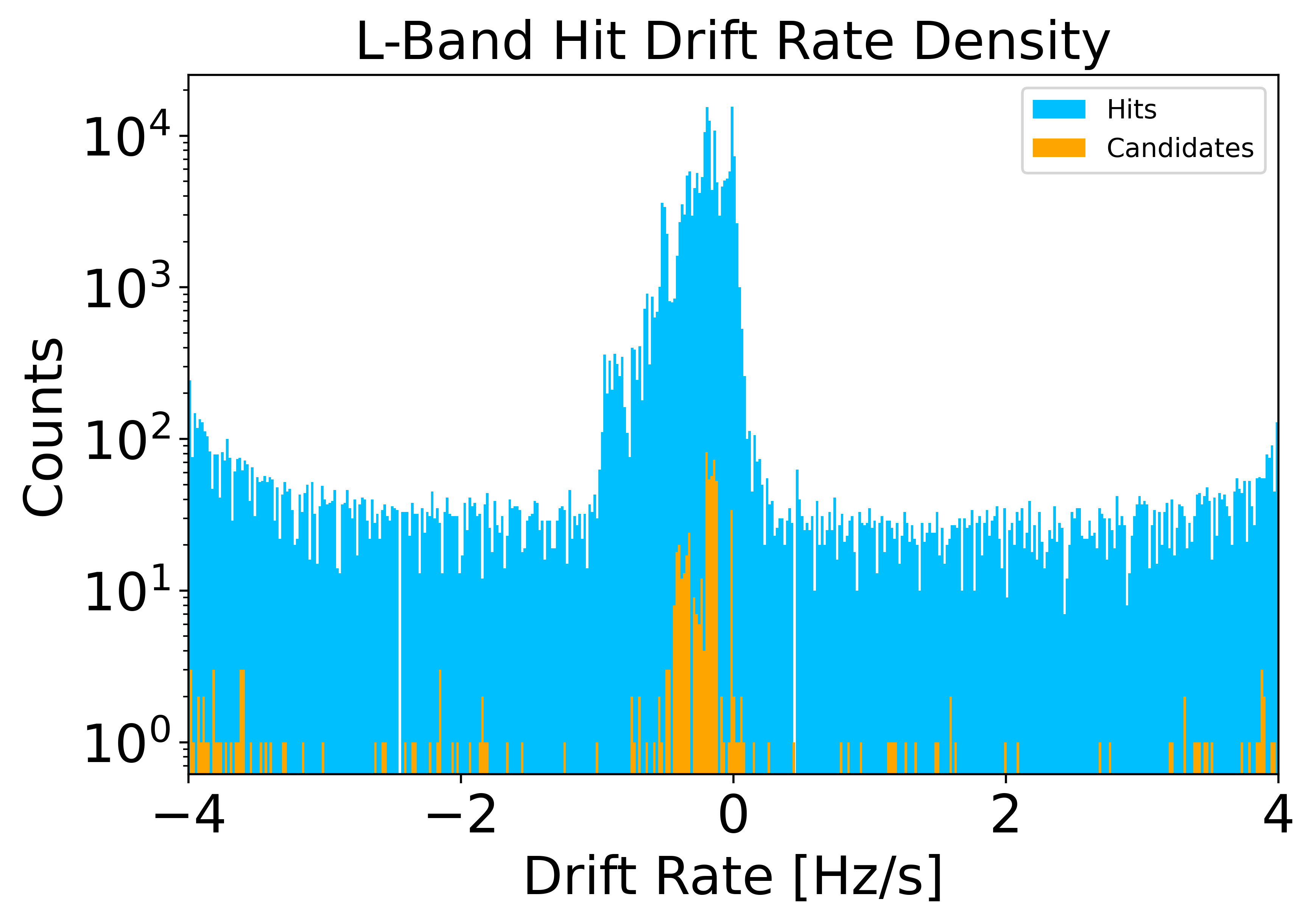}%
  \label{fig:drift-density-L}%
}\qquad
&
\subfloat[\snr\ hit density, L-band]{%
  \includegraphics[width=0.7\columnwidth]{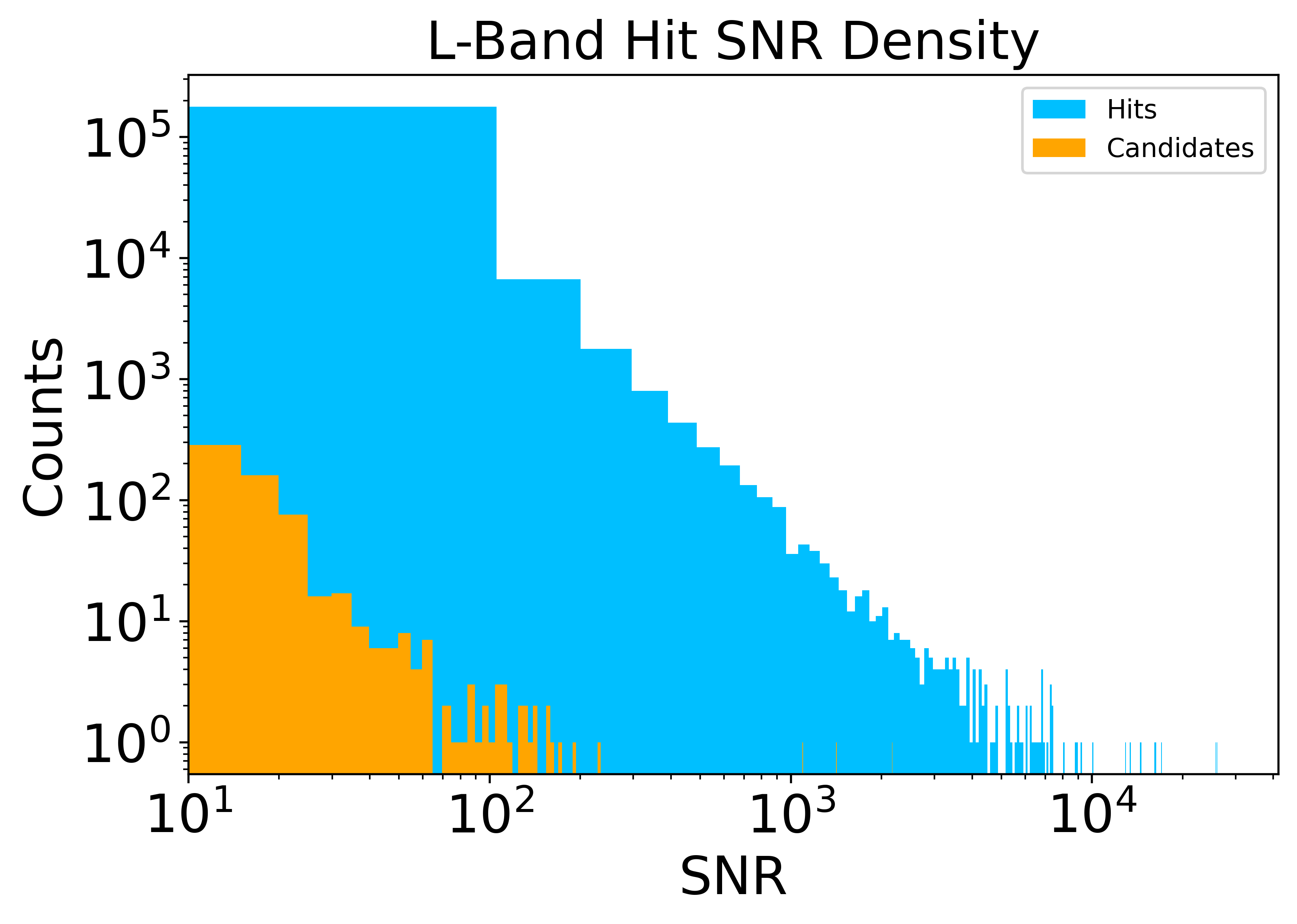}
  \label{fig:snr-density-L}%
}\qquad
&
\subfloat[Frequency hit density, L-band]{%
  \includegraphics[width=0.7\columnwidth]{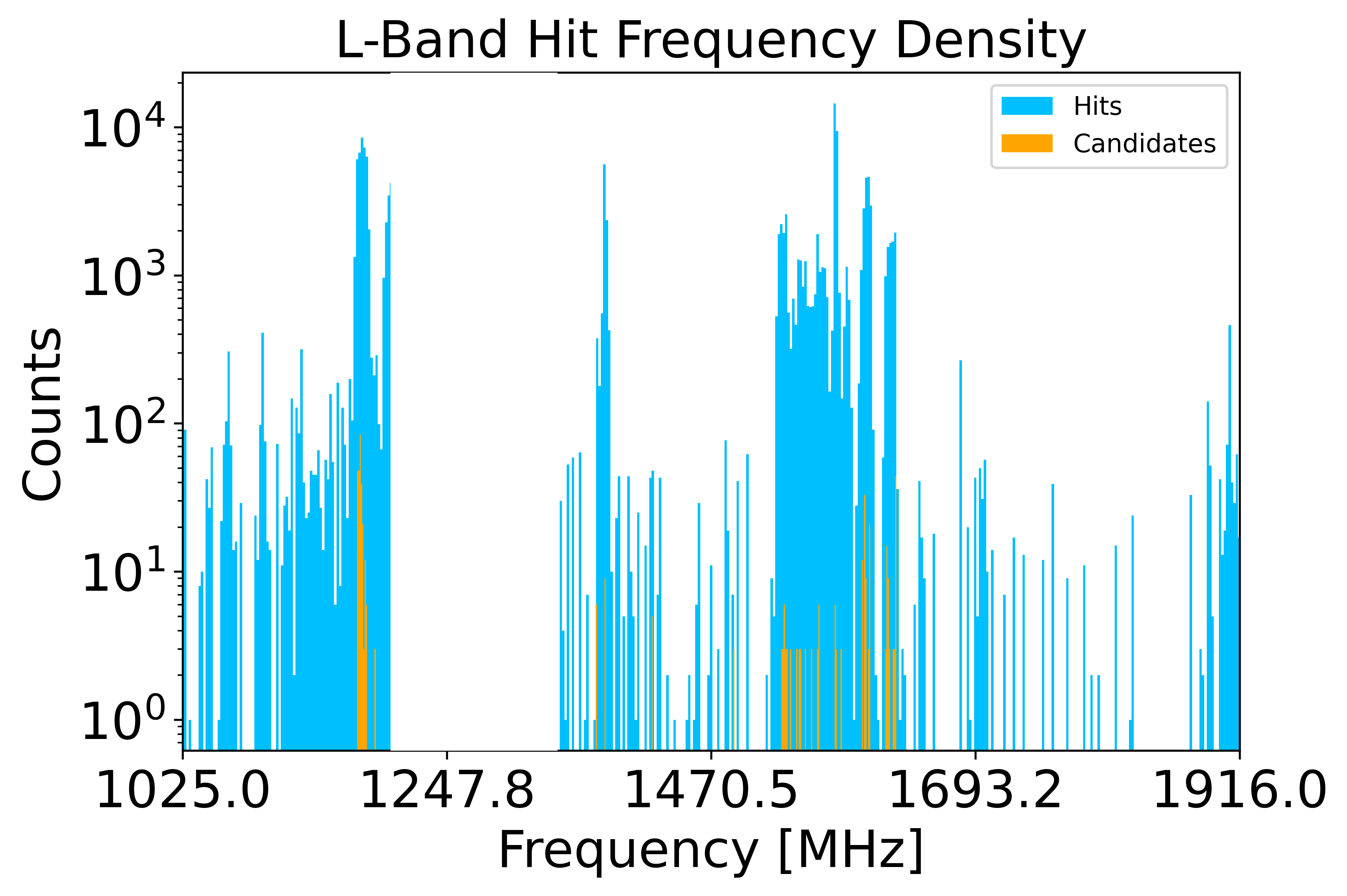}
  \label{fig:candidate-freq-density-lband}
}\qquad
\\
\subfloat[Drift rate hit density, S-band]{%
  \includegraphics[width=0.7\columnwidth]{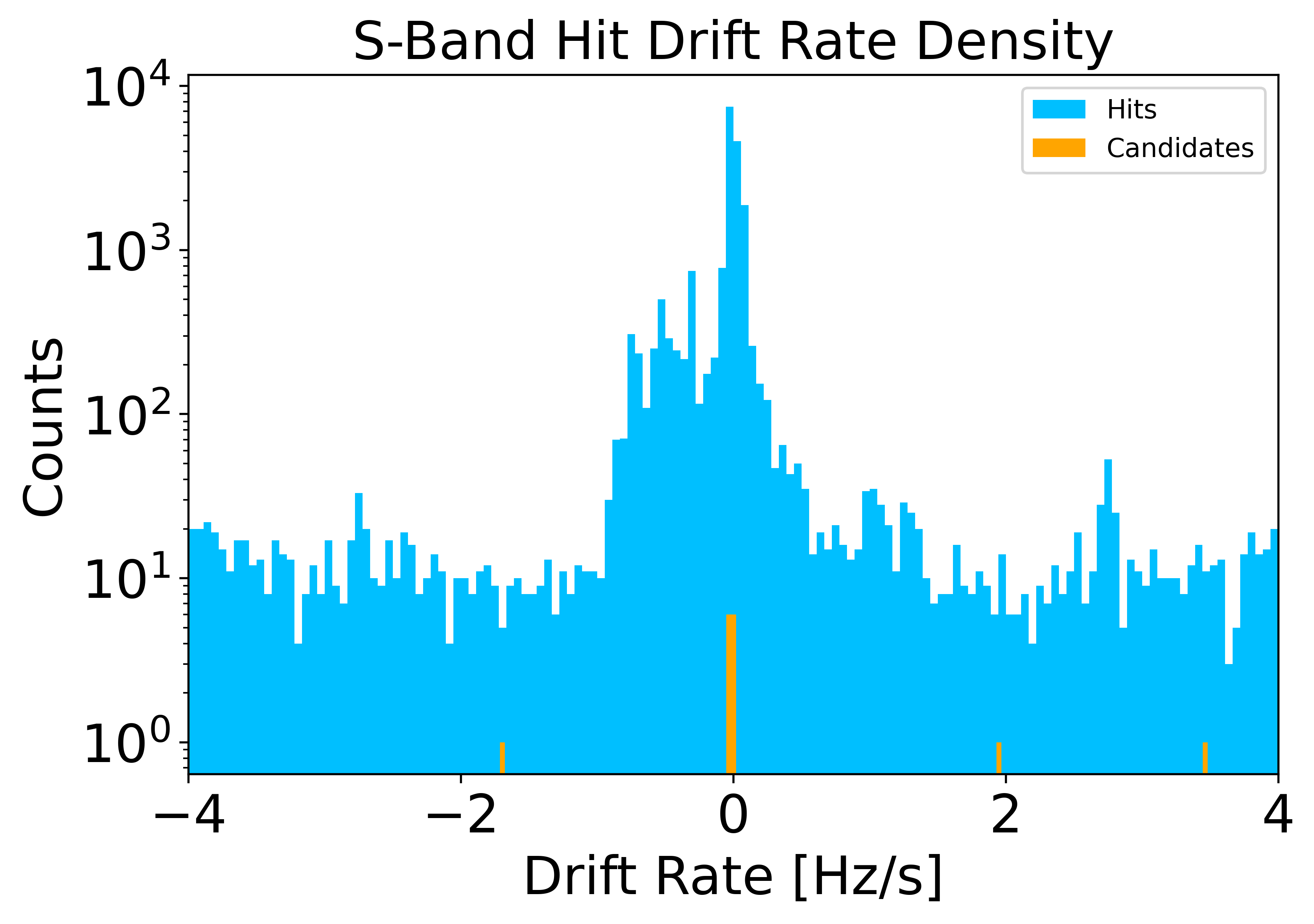}%
  \label{fig:drift-density-S}%
}\qquad
&
\subfloat[\snr\ hit density, S-band]{%
  \includegraphics[width=0.7\columnwidth]{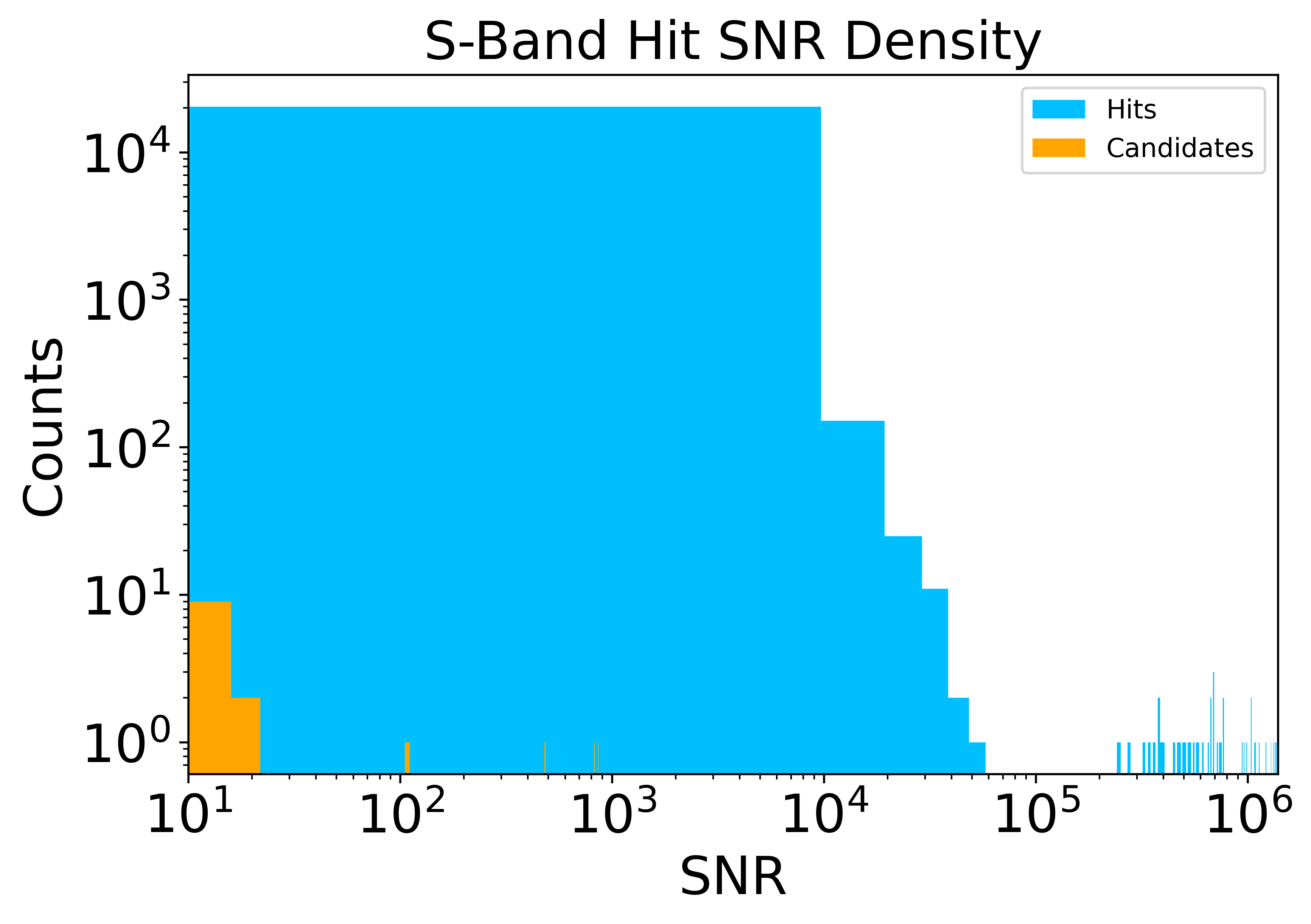}%
  \label{fig:snr-density-S}%
}\qquad
&
\subfloat[Frequency hit density, S-band]{%
  \includegraphics[width=0.7\columnwidth]{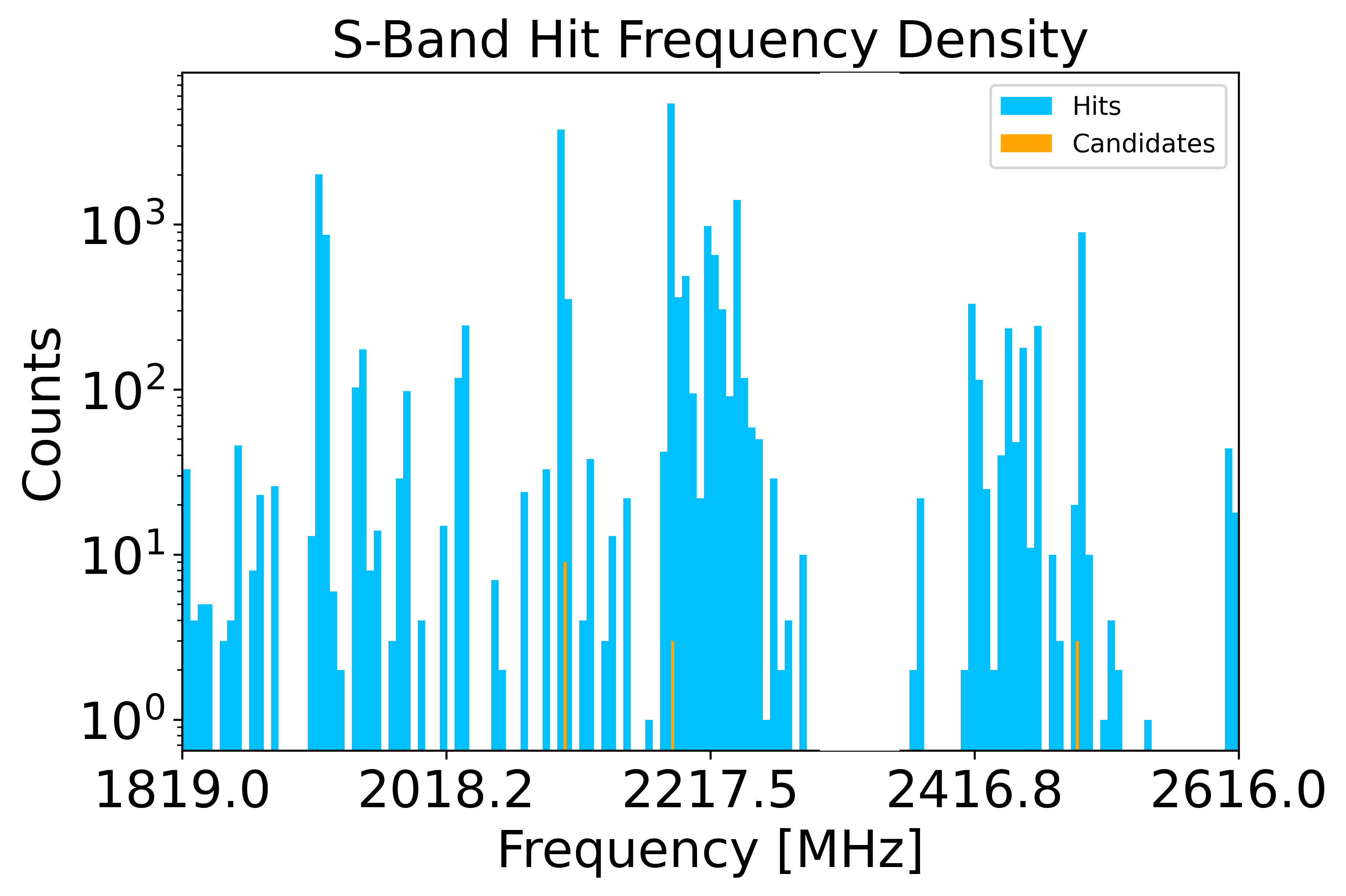}
  \label{fig:candidate-freq-density-sband}
}\qquad
\\
\subfloat[Drift rate hit density, C-band]{%
  \includegraphics[width=0.7\columnwidth]{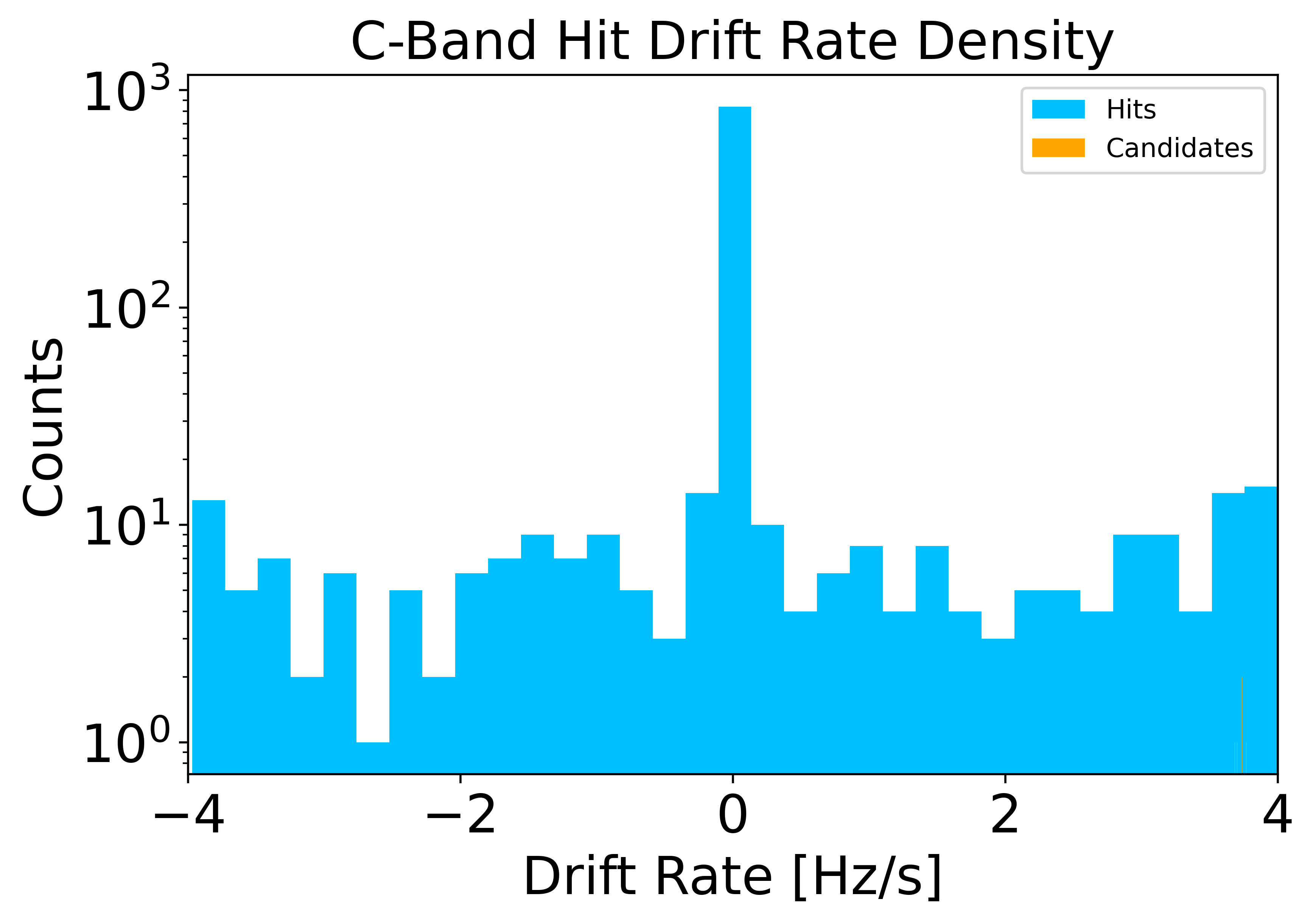}%
  \label{fig:drift-density-S}%
}\qquad
&
\subfloat[\snr\ hit density, C-band]{%
  \includegraphics[width=0.7\columnwidth]{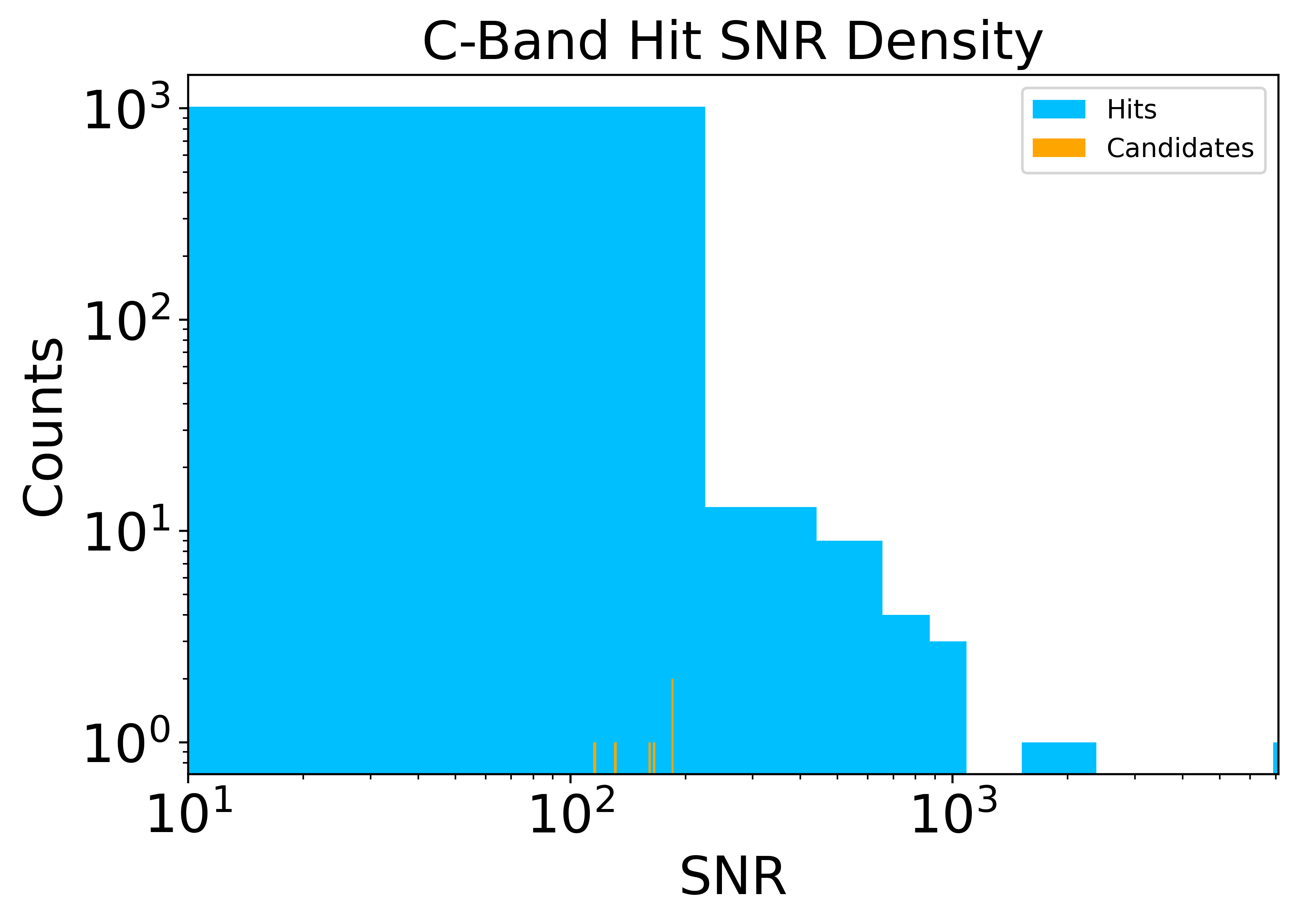}
  \label{fig:snr-density-C}%
}\qquad
&
\subfloat[Frequency hit density, C-band]{%
  \includegraphics[width=0.7\columnwidth]{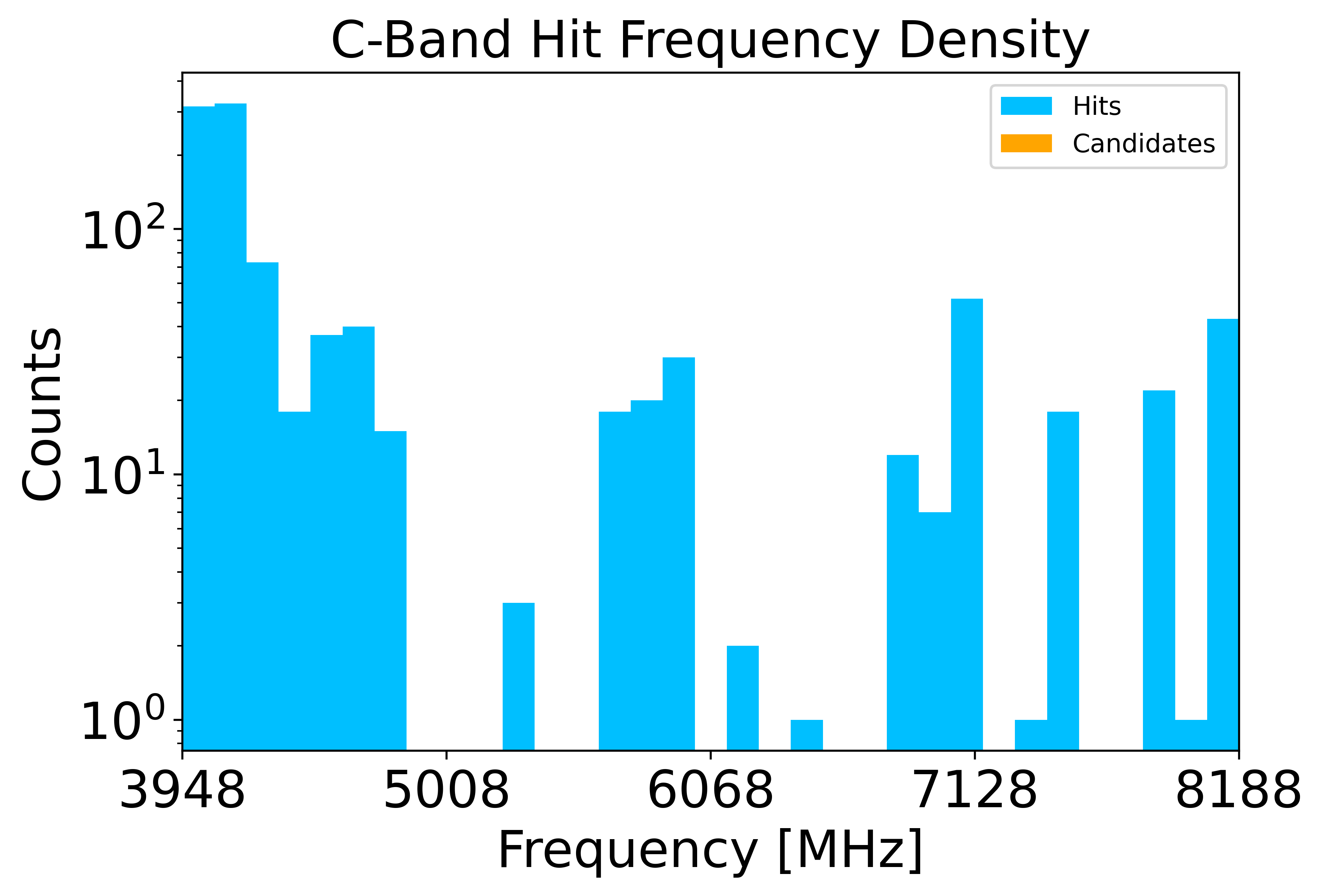}
  \label{fig:candidate-freq-density-cband}
}\qquad
\\
\subfloat[Drift rate hit density, X-band]{%
  \includegraphics[width=0.7\columnwidth]{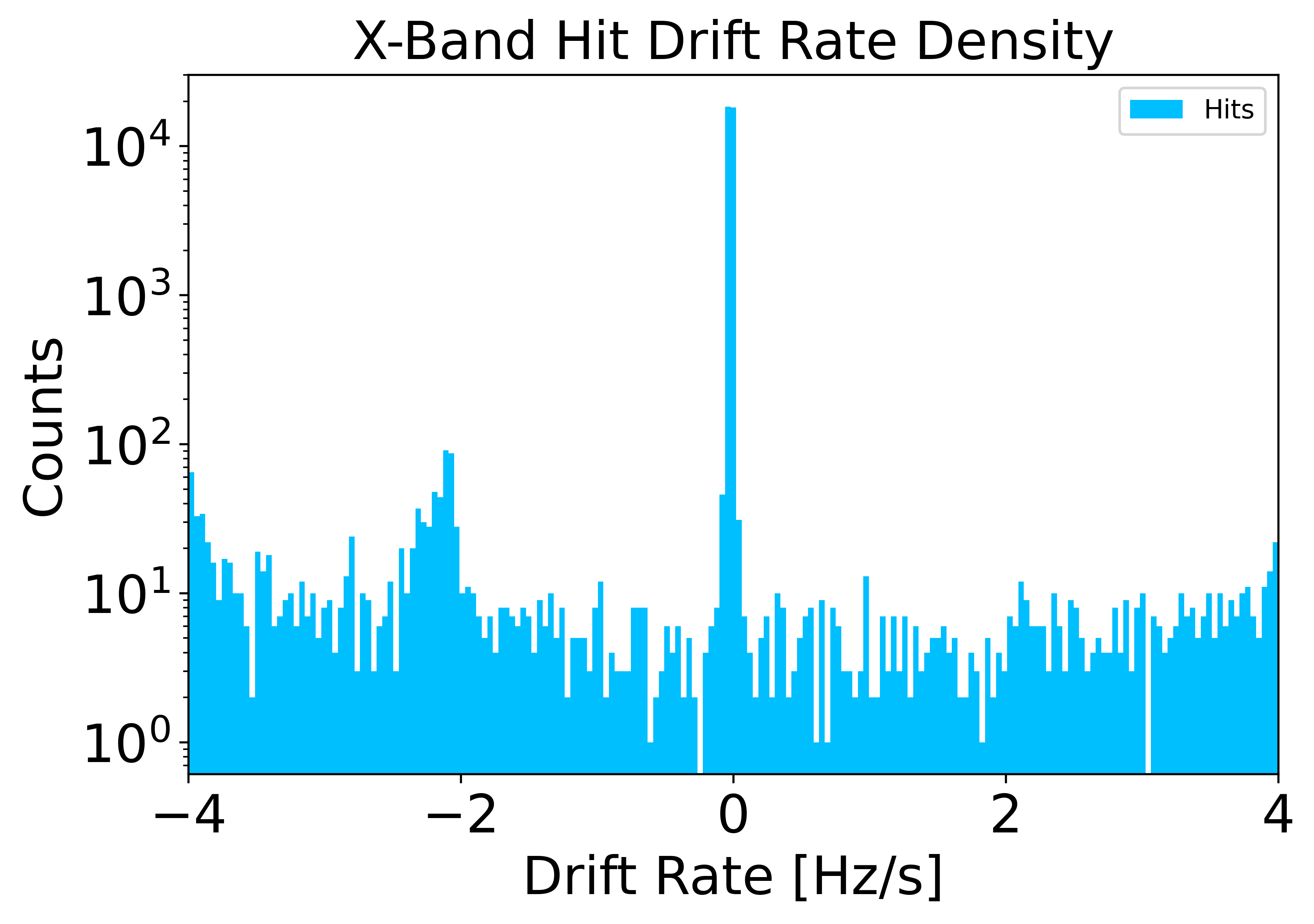}%
  \label{fig:drift-density-X}%
}\qquad
&
\subfloat[\snr\ hit density, X-band]{%
  \includegraphics[width=0.7\columnwidth]{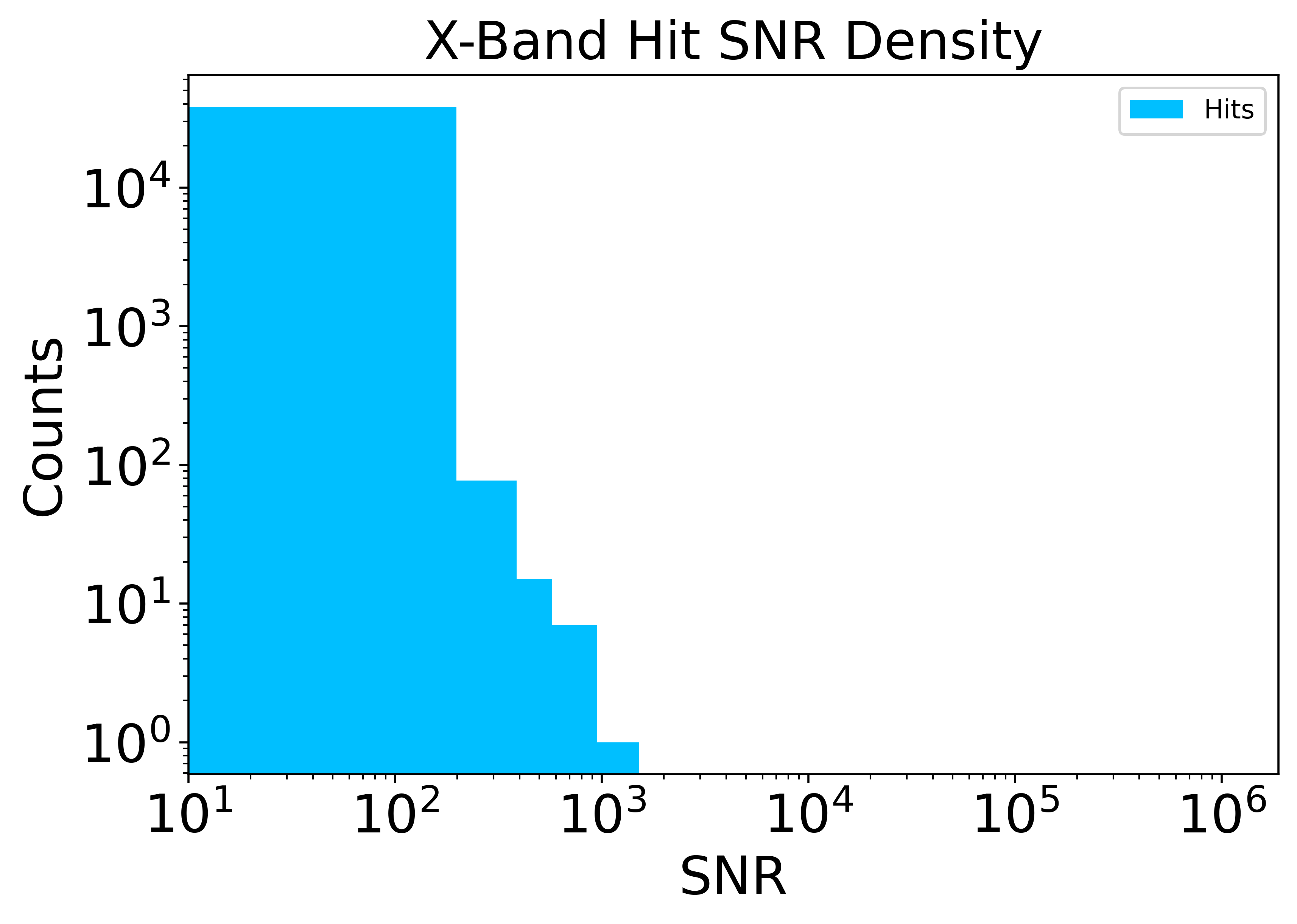}%
  \label{fig:snr-density-X}%
}\qquad
&
\subfloat[Frequency hit density, X-band]{%
  \includegraphics[width=0.7\columnwidth]{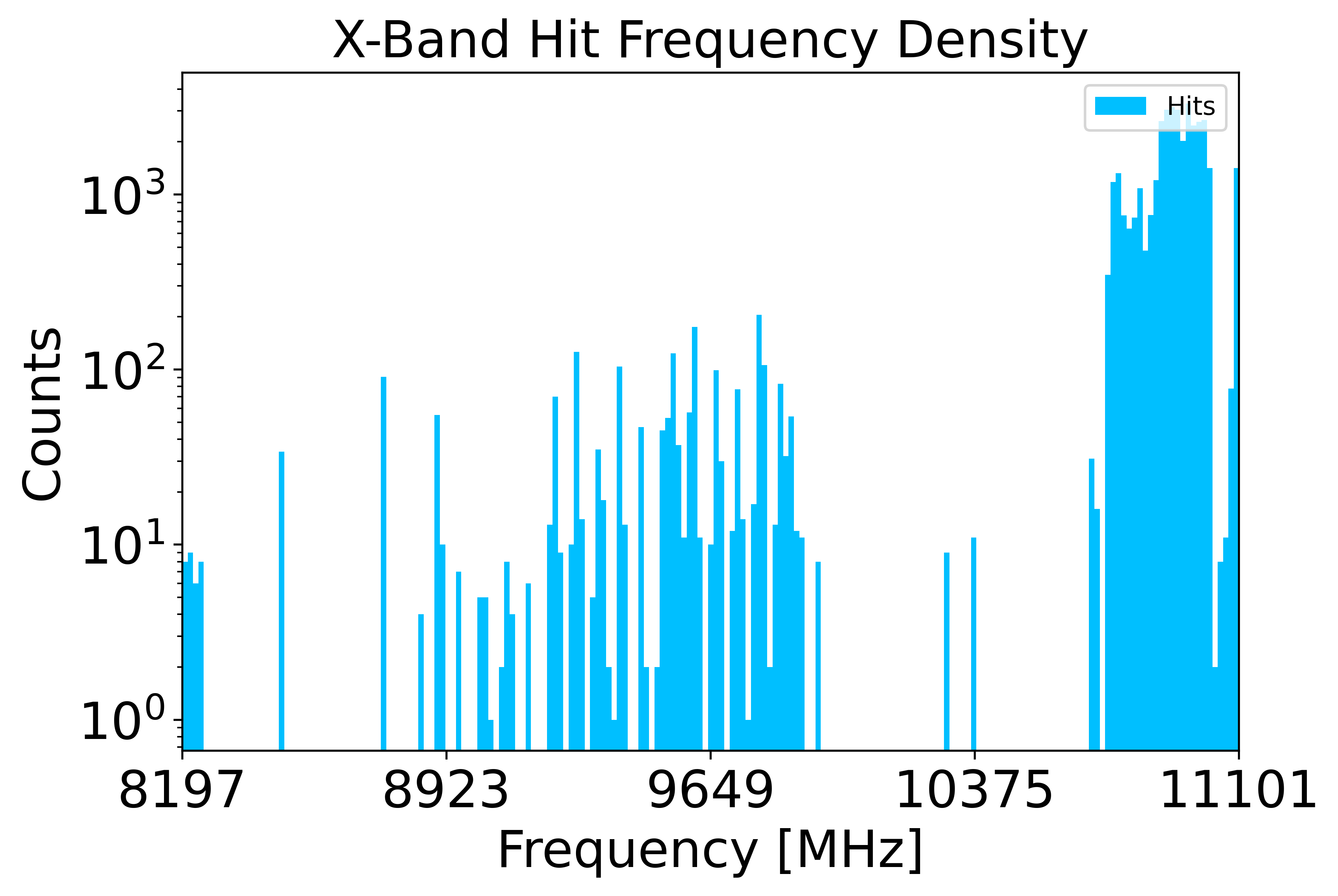}
  \label{fig:candidate-freq-density-xband}
}\qquad
\\
\end{tabular}
\caption{Histograms of the drift rate, \snr\ , and frequency density of hits (blue) and candidates (orange).  Note \reffig{fig:drift-density-X}, \reffig{fig:snr-density-X}, and \reffig{fig:candidate-freq-density-xband} reflect no candidates being found at X-band.\label{fig:drift-snr-density-histograms}}
\end{figure*}

\begin{table*}[t!]\label{tab:observation-summary}
\begin{center}
\caption{\label{tab:candidates_by_band} Candidates found across all bands.
}
\begin{tabular}{lccccccc}
\hline
Receiver & No. cadences & Hours & No. hits & No. events & No. candidates & TFM & Transmitter Limit\footnote{The limit on the existence of putative transmitters.  Indicates the percentage of observed stars which possess transmitters above the $\text{EIRP}_\text{min}$ threshold.  That is, of the stars observed at a particular band, the transmitter limit is the maximum proportion of those which possess narrowband transmitters.  Values computed using a one-sided Poisson confidence interval, given a $50\%$ probability of observing a signal if present.}\\
\hline
\hline
L        & \ncadencegbtL  &  \nhrgbtL  & \nhitsL   & \neventsL&\ncandL & \tfmLcad & \ptransL\\%
S        & \ncadencegbtS  &  \nhrgbtS  & \nhitsS   & \neventsS&\ncandS & \tfmScad & \ptransS\\%
C        & \ncadencegbtC  &  \nhrgbtC  & \nhitsC   & \neventsC& \ncandC & \tfmCcad & \ptransC\\
X        & \ncadencegbtX\footnote{There were two cadences of observations performed for TIC232967666.}  &  \nhrgbtX  & \nhitsX   &\neventsX & \ncandx& \tfmXcad & \ptransX\\
\hline
Total    & \ncadencegbtsum&   \nhrgbtsum& \nhitssum & \neventssum & \ncand & \tfmCombined & \ptrans \\
\hline
\end{tabular}
\end{center}
\end{table*}
\begin{figure*}[b!]
\begin{center}
\includegraphics[width=0.75\linewidth]{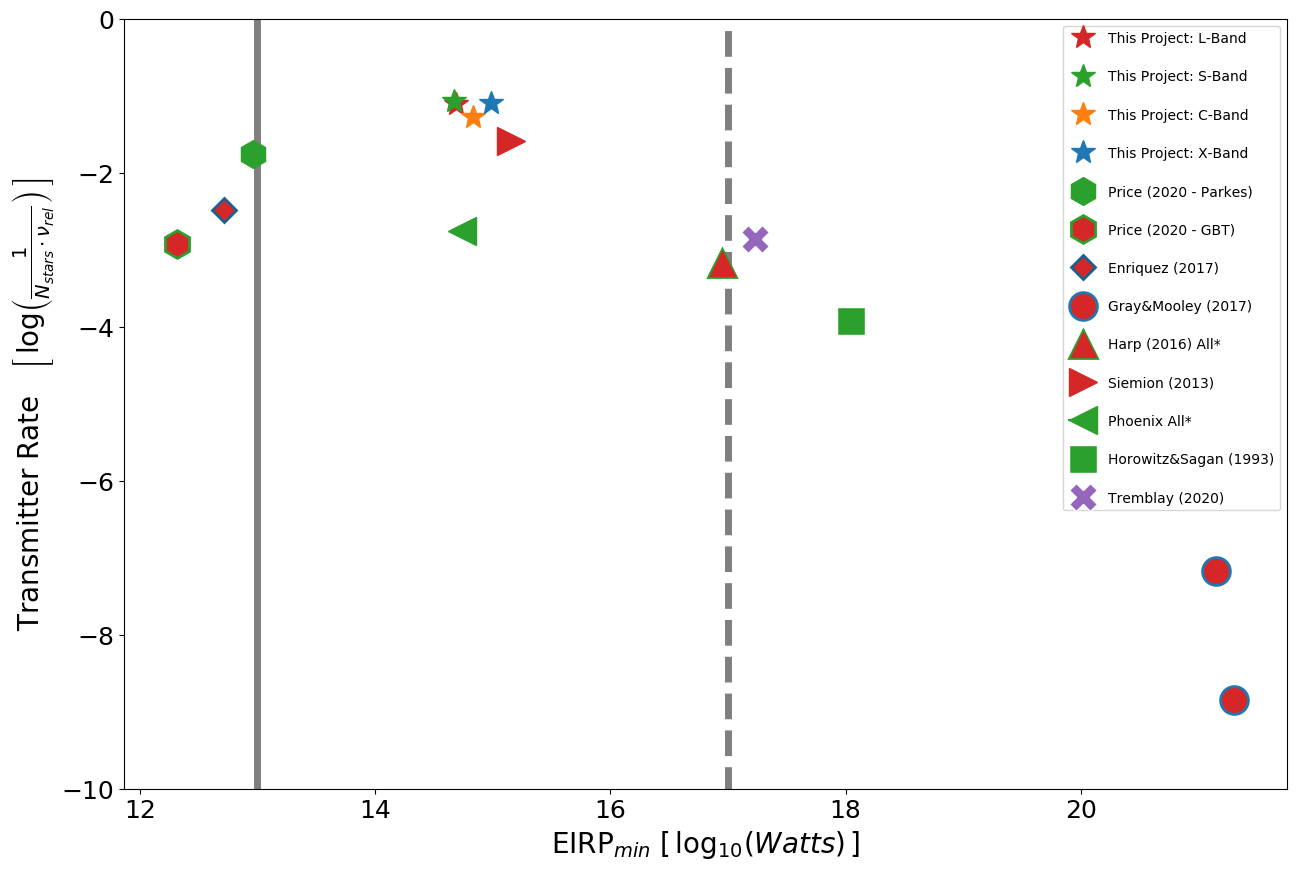}
\protect\caption{A comparison of the derived transmitter rate limits of this work with previous surveys, where red, green, orange, and blue correspond to searches conducted at L-, S-, C-, and X-band respectively.  Respectively, the vertical solid and dashed lines represent the EIRP of the Arecibo planetary radar, and the total solar power incident on Earth.  Note that the \citet{Tremblay:2020} data point represents a survey conducted at 113 MHz with the Murchison Widefield Array (MWA).  The values for Project Phoenix are from \citet{Price:2020}.  The \citet{Horowitz:1993} data point represents an all-sky search near the 1420 MHz neutral hydrogen line, while \citet{Gray:2017dr} searched the same line with the Very Large Array (VLA).  \citet{Harp:2016} conducted a search from 1-9 GHz with the Allen Telescope Array (ATA).\label{fig:seti-comparison}
}
\end{center}
\end{figure*}

Narrowband drifting signals with \snr\ $> 10$ detected by \tseti\ are referred to as `hits'.  A set of hits present in all `ON' observations falling within the range of frequencies subtended by a signal with constant drift are grouped into an `event'.  That is, the $n$th event $\displaystyle \varepsilon_{n}$ can be defined by Eq.~\ref{eqn:event-definition}, where $\displaystyle t_{\text{obs}}$ is the duration of each observation (5\,min), $\displaystyle \nu_{h}$ is the frequency of a hit $h$ present in the $k$th `ON' observation, and $\displaystyle \nu_{0}$ and $\displaystyle \dot{\nu_{0}}$ are the frequency and drift rate, respectively, of a hit present in the first `ON' observation $A_{0}$ in the cadence ABACAD or ABABAB.  Events which contain no hits present in all `OFF' observations are defined as `potential candidates'.  Thus, the $n$th potential candidate $\displaystyle \rho_{n}$ can be defined by Eq.~\ref{eqn:candidate-definition}.

\begin{widetext}
\begin{equation}\label{eqn:event-definition}
\textstyle \varepsilon_{n} = \left\{h \in A_{k} \::\: \nu_{0} - \dot{\nu_{0}} \cdot t_{\text{obs}}\; \leq \;\nu_{h} \;\leq\; \nu_{0} + \dot{\nu_{0}} \cdot t_{\text{obs}} \right\}
\end{equation}

\begin{equation}\label{eqn:candidate-definition}
\textstyle \rho_{n} = \left\{h \in A_{k} \::\: h \not\in B \wedge \nu_{0} - \dot{\nu_{0}} \cdot t_{\text{obs}}\; \leq \;\nu_{h} \;\leq\; \nu_{0} + \dot{\nu_{0}} \cdot t_{\text{obs}} \right\}
\end{equation}
\end{widetext}



\subsection{Signal Distribution}\label{hit-distributions}

Events and candidates were found using the \tseti\ \texttt{find\_event} method for all full cadences, which found a total of $\neventssum$ events and $\ncand$ candidates across all four bands.  A breakdown of the results is summarized in \reftab{tab:observation-summary}.  The drift rate, \snr, and frequency distribution of hits and candidates detected at each band is shown in \reffig{fig:drift-snr-density-histograms}.  A sample of plots of candidate events is included in Appendix \ref{sec:best-candidates}.  

L-band targets accounted for the largest share of hits ($76\%$), while X-band amounts to the largest portion of hits among the remaining bands ($15\%$), with S- and C-band accounting for $8\%$ and $1\%$ of total hits, respectively.  We find that L-band accounts for $80\%$ of total events, a slight increase compared to its hits proportion.  Despite S-band comprising fewer hits than X-band, more events were recorded at S-band ($13\%$) than X-band ($6\%$), whereas the relative proportion of C-band events remained unchanged throughout ($1\%$).  Overall, $17\%$ of the total hits were present in candidate events -- the hit-event `conversion' rate\footnote{In general, the $\displaystyle \text{A-B}$ conversion rate is $\displaystyle \text{A $\xrightarrow{}$ B} = \frac{\text{total A}}{\text{total B}}$, and, in context, roughly describes how much of $\displaystyle \text{A}$ is contained in $\displaystyle \text{B}$}.  The L-band hit-event conversion rate was $18\%$, a low figure compared to 
those seen at S- ($27\%$) and C-band ($26\%$), which may indicate a stronger RFI presence.  The proportion of candidates found at L-band was significantly higher than was previously for hits, accounting for $97\%$ of total candidates.  \reftab{tab:summary-stats} gives an overview of these figures.  

\begin{table}[!htb]
\captionsetup{size=footnotesize}
\caption{Event Groups} \label{tab:event-groups}
\setlength\tabcolsep{0pt} 
\smallskip 
\begin{tabular*}{\columnwidth}{@{\extracolsep{\fill}}lccc}
\hline
Receiver & $\displaystyle N_\text{event groups}$ & $\displaystyle N_\text{single sources}$ & $\displaystyle N_\text{non-RFI}$    \\
\hline
\hline
L-Band  &   \nevgL  &   \nssL   &   0\\
S-Band  &   \nevgS  &   \nssS   &   0\\
C-Band  &   \nevgC  &   \nssC   &   0\\
X-Band  &   \nevgX  &   \nssX   &   0\\
\hline
\end{tabular*}
\end{table}

\subsection{Event Groups}
Following \citet{Price:2020}, events were grouped into ``event groups'' of 125\,kHz frequency bins, in which the spacing between the highest and lowest start frequencies $\Delta\nu_{\text{event}}$ in each bin was computed.  From this value, the central frequency $\displaystyle \nu_{\text{event}}$ for each event group was computed.  The larger the cluster size, the greater the RFI (or technosignature) presence.  

The candidate events at L-band were grouped into 135 frequency bins of 125\,kHz, the lowest grouping at 1172\,MHz, and the highest at 1626\,MHz.  

The statistical variance (and thus, standard deviation $\textstyle \sigma$) in frequency, \snr, and drift rate within an event group were found for each event group.  A graphical depiction is shown in \reffig{fig:lband-event-groups-variance}.  We define single sources of interference or technosignature as event groups whose events demonstrate uniformity in frequency, \snr, and drift rate, because signals which share similar characteristics are more likely to have a common origin.  As such, we select event groups with $\displaystyle \sigma_{\nu} < 1$, $\displaystyle \sigma_{\texttt{S/N}} < 1$, and $\displaystyle \sigma_{\dot{\nu}} < 1$ as single sources.  The results are shown in \reftab{tab:event-groups}.  Examples of single sources are \reffig{fig:single-source-1} and \reffig{fig:single-source-2}.

The \bl \blimpy package \citep{ascl:blimpy} was used to produce diagnostic waterfall plots of the 20 single sources, in which upon visual inspection, interesting sources traceable to known RFI can be identified.  A sample of these plots are included in Appendix \ref{sec:appendix-event-groups}.  All single-source event groups can be traced back to observations of 5 unique targets.

\subsection{Transmitter Rate Limit}\label{sec:transmitter-rate}
We can calculate the percentage of stars which could possess high duty cycle transmitters above the $\text{EIRP}_\text{min}$ threshold, given our lack of detections.  That is, of the stars observed at a particular band, the transmitter limit is the maximum proportion of those which possess narrowband transmitters.  Values are computed using a one-sided Poisson confidence interval, assuming a conservative $50\%$ probability of observing a signal if present.

The transmitter rate limit was calculated at L-, S-, C-, and X-band, the results of which are presented in \reftab{tab:candidates_by_band}.  Our transmitter limit of $12.72\%$ indicates that fewer than $12.72\%$ of the observed stars have putative transmitters operating in the range of 1 -- 11 GHz.

A comparison of the derived transmitter rate limits to those of previous surveys is shown in \reffig{fig:seti-comparison}, where red, green, orange, and blue correspond to searches conducted at L-, S-, C-, and X-band respectively. Our study provides similar constraints to previous studies, but over a much wider range of frequencies.  

Although we calculate transmitter rates solely considering the TESS targets, the dataset also includes significant ``bycatch" of other stars (and, indeed, background galaxies) in the GBT pointings \citep[see][]{Wlodarczyk_Sroka_2020}. Hence additional constraints on the prevalence of more luminous transmitters at larger distances from Earth could in principle be calculated, in addition to those for the TESS targets presented here.

\section{Conclusions}\label{sec:conclusions}

We report on the Breakthrough Listen technosignature search of 28 stellar targets selected from the \tess\ Input Catalog identified as potential exoplanet hosts.  Our observations spanning 1 -- 11 GHz were searched for narrowband signals exhibiting drift rates within $\pm 4$\,Hz\,s$^{-1}$ above a minimum \snr\ threshold of 10, and no candidate signals unattributable to RFI were found.  We derive an $\text{EIRP}$ threshold of \eirplimit\ and establish some of the deepest limits to date over such a wide band.

\subsection{Future Work}\label{sec:future}

Breakthrough Listen is continuing to observe a larger sample of TOI targets.  Furthermore, \tess\ has begun its extended mission and so will continue to refine its TOI catalog\footnote{\url{https://tess.mit.edu/observations/}}, enabling future SETI searches to have a larger number of confirmed, nearby exoplanets for study.

\section{Acknowledgements}

The Breakthrough Prize Foundation funds the Breakthrough Initiatives which manages $\bl$.  The Green Bank Observatory facility is supported by the National Science Foundation, and is operated by Associated Universities, Inc. under a cooperative agreement.  We thank the staff at Green Bank Observatory for their support with operations.  Raffy Traas was funded as a participant in the Berkeley SETI Research Center Research Experience for Undergraduates Site, supported by the National Science Foundation under Grant No.~1950897.

We thank Richard Elkins and Luigi Cruz for help with development and debugging of turboSETI.  We thank the referee for their helpful comments.

\software{\tseti \citep{Enriquez:2017, tseti_paper}, \blimpy \citep{ascl:blimpy}, \textsc{NumPy} \citep{numpy:2020}, \textsc{pandas} \citep{pandas_software, pandas_paper}, \textsc{astropy} \citep{astropy:2013, astropy:2018}, \textsc{Docker} \citep{docker}}

\nocite{Harp:2016, Horowitz:1993, Gray:2017dr}
\bibliographystyle{aasjournal}
\bibliography{ms}

\appendix
\section{\tseti\ in Google Cloud Platform} \label{sec:appendix-gcp}
This work is the result of an endeavor to migrate the \tseti pipeline onto the GCP.  GCP Compute Engine instances were used to conduct the analysis instead of the compute nodes at the UC Berkeley Data Center.  Each of the 20 instances was mounted to a GCS Bucket to access the fine-frequency resolution data products and installed with a containerized version of \tseti\ as a \dock\ image to search for narrowband drifting signals.  Five Compute Engines were assigned to analyze the data at each of the bands (L, S, C, X).  One of these instances acted as the ``head node'' and provided each Compute Engine a unique list of files to analyze.  The head node then collected and searched the \tseti\ raw output files for events.  Events which passed \tseti\ inspection were plotted.
\begin{figure*}[!ht]
\begin{center}
\includegraphics[width=\linewidth]{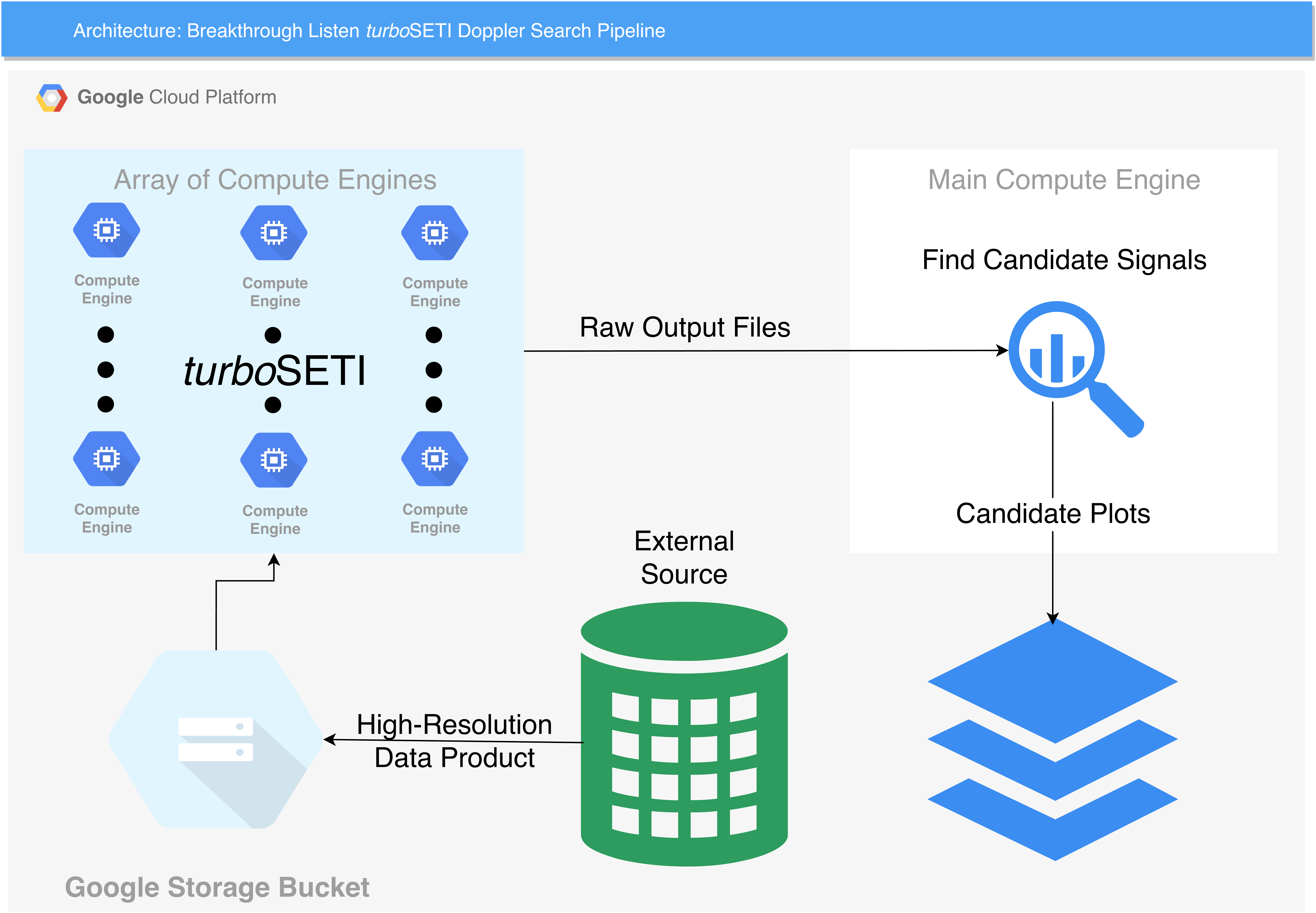}
\caption{The \tseti\ pipeline adapted for compatibility with Google Cloud Platform\label{fig:turbo-cloud-architecture}
}
\end{center}
\end{figure*}

\pagebreak

\section{Targets}\label{sec:targets}
\reftab{tab:bl-tess-targets} presents the 28 targets studied in this work, which have been observed with the GBT at all four L, S, C, and X bands.  Each of the targets is present in the TOI catalog.  Target data is sourced from the ExoFOP-TESS.
\begin{table*}[h]\label{tab:bl-tess-targets}
\begin{center}
\caption{\label{tab:target-details} \textbf{Targets}}

\begin{tabular}{lcllcr}
\hline
TIC ID & TOI ID & RA (J2000) &  Dec (J2000) & Distance (pc)\\
\hline
\hline
154089169 & 1174.01 & 13:56:52.12 & 68:37:05.56 &  95\\
154840461 & 1153.01 & 12:10:45.97 & 85:42:18.28 & 147\\
154872375 & 1135.01 & 12:53:35.13 & 85:07:46.18 &  114\\
158002130 & 1180.01 & 14:18:13.39 & 82:11:37.55 & 72\\
159510109 & 1141.01 & 16:30:08.33 & 80:18:27.96 & 97\\
160268701 & 1143.01 & 12:10:09.28 & 77:21:08.15 & 192\\
198213332 & 1131.01 & 16:33:44.05 & 61:43:06.00 & 223\\
230088370 & 1176.01 & 16:29:25.60 & 71:30:21.14 &  296\\
232967666 & 1167.01 & 13:12:36.57 & 71:37:05.30 &  210\\
233087860 & 1184.01 & 18:08:49.08 & 60:40:43.62 &  59\\
266500992 & 1655.01 & 04:04:41.34 & 52:15:25.17 & 165\\
267489265 & 1132.01 & 19:55:07.51 & 54:46:53.20 & 287\\
267694283 & 1656.01 & 04:12:55.20 & 53:41:13.38 & 106\\
281731203 & 685.01 & 10:52:07.75 & 00:29:35.40 &  213\\
284450803 & 1142.01 & 19:22:50.15 & 56:36:55.39 &  299\\
287196418 & 1190.01 & 19:30:56.19 & 59:24:13.87 &  281\\
288185138 & 1200.01 & 14:51:21.94 & 82:57:08.13 & 224\\
289539327 & 1186.01 & 16:46:32.03 & 65:42:54.13 & 295\\
294176967 & 1170.01 & 20:03:13.23 & 52:02:27.03 & 880\\
302518439 & 1169.01 & 08:56:28.81 & 72:31:59.79 & 262\\
320525204 & 1140.01 & 17:39:50.77 & 56:04:44.30 & 123\\
341544930 & 1182.01 & 13:26:27.52 & 65:18:22.84 & ---\\
349827430 & 1148.01 & 10:47:38.17 & 71:39:20.62 &  97\\
359496368 & 1178.01 & 18:36:24.25 & 55:23:30.41 &  37\\
372757221 & 1187.01 & 10:51:34.36 & 81:19:19.83 & 221\\
458478250 & 1165.01 & 15:28:35.19 & 66:21:31.35 & 126\\
459970307 & 1154.01 & 16:59:41.77 & 64:41:57.17 & 94\\
470315428 & 1673.01 & 04:18:35.64 & 52:51:54.12 & 472\\
\hline
\end{tabular}
\end{center}
\end{table*}


\clearpage
\section{Examples of Candidate Events}\label{sec:best-candidates}
There were \ncand\ candidate events detected by \tseti\ across all four bands.  \reffig{fig:best-candidates} shows a candidate event found at each band.  These events passed \tseti\ inspection as having signals with $\snr\ > 10$ present in the `ON' observations and none in the `OFF' observations.  By visual inspection, it is easily verifiable that there are emissions of $\snr\ < 10$ present in `OFF' observations--characteristic of RFI.  In this way, each of the \ncand\ candidate events were rejected.
\begin{figure*}[hb]
\centering 
\subfloat[L-Band]{%
  \includegraphics[width=.42\columnwidth]{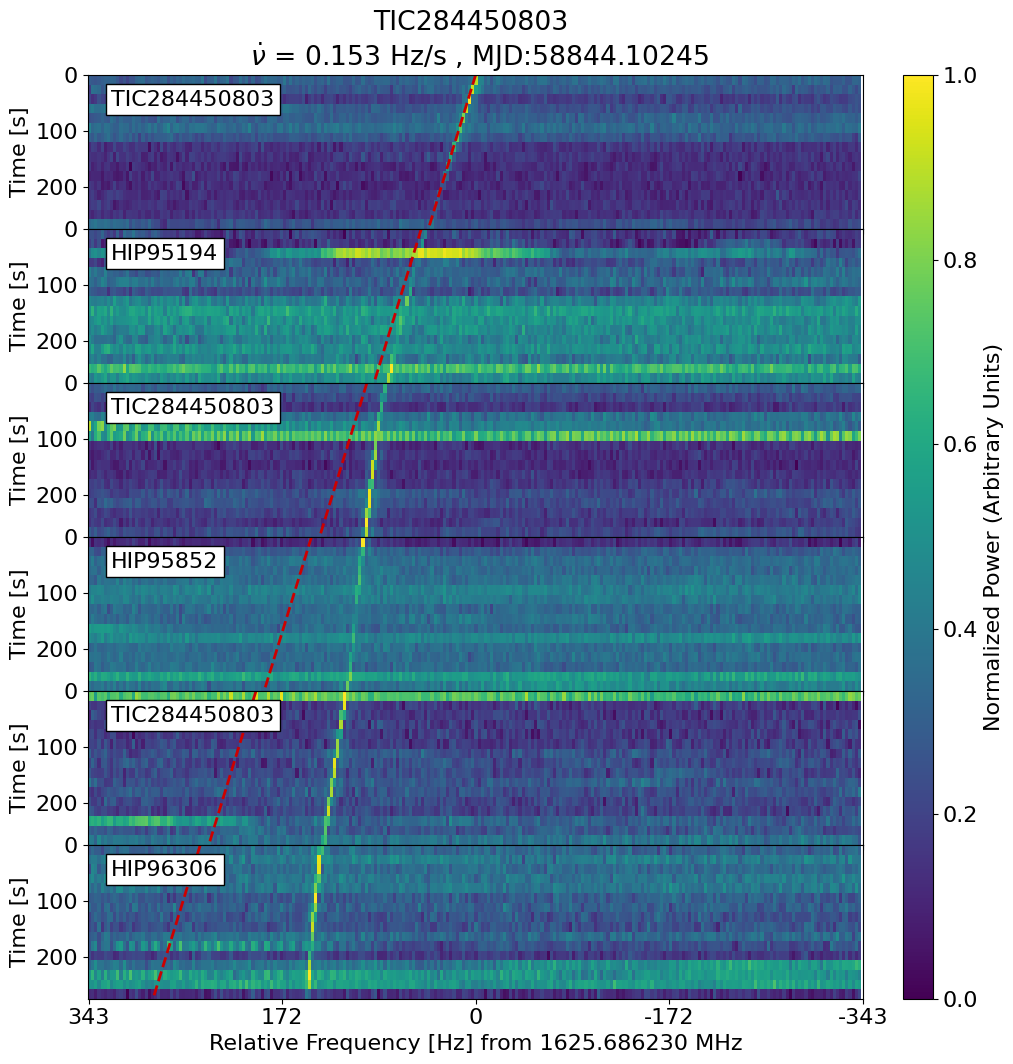}
  \label{fig:best-candidate-LBand}%
}\qquad
\subfloat[S-Band]{%
  \includegraphics[width=.42\columnwidth]{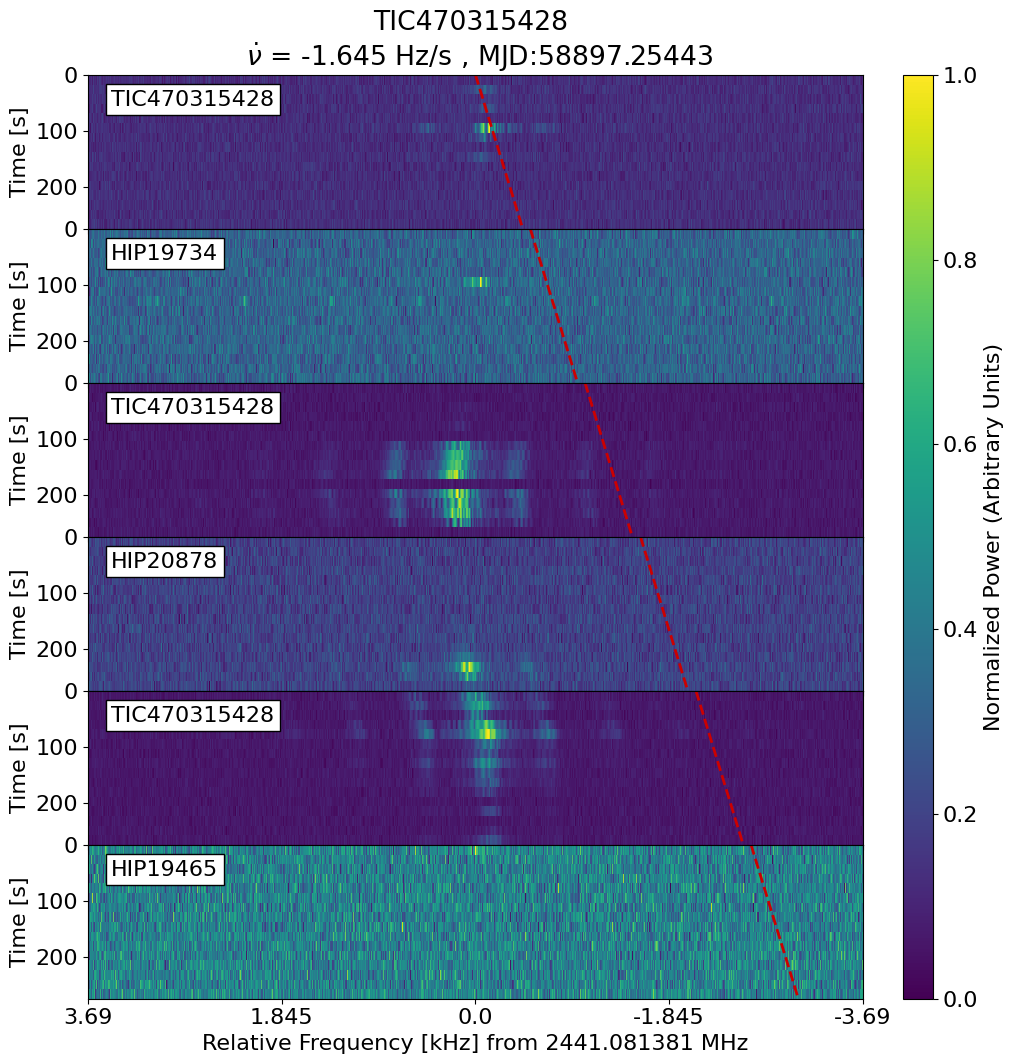}
  \label{fig:best-candidate-SBand}%
}\qquad
\subfloat[C-Band]{%
  \includegraphics[width=.42\columnwidth]{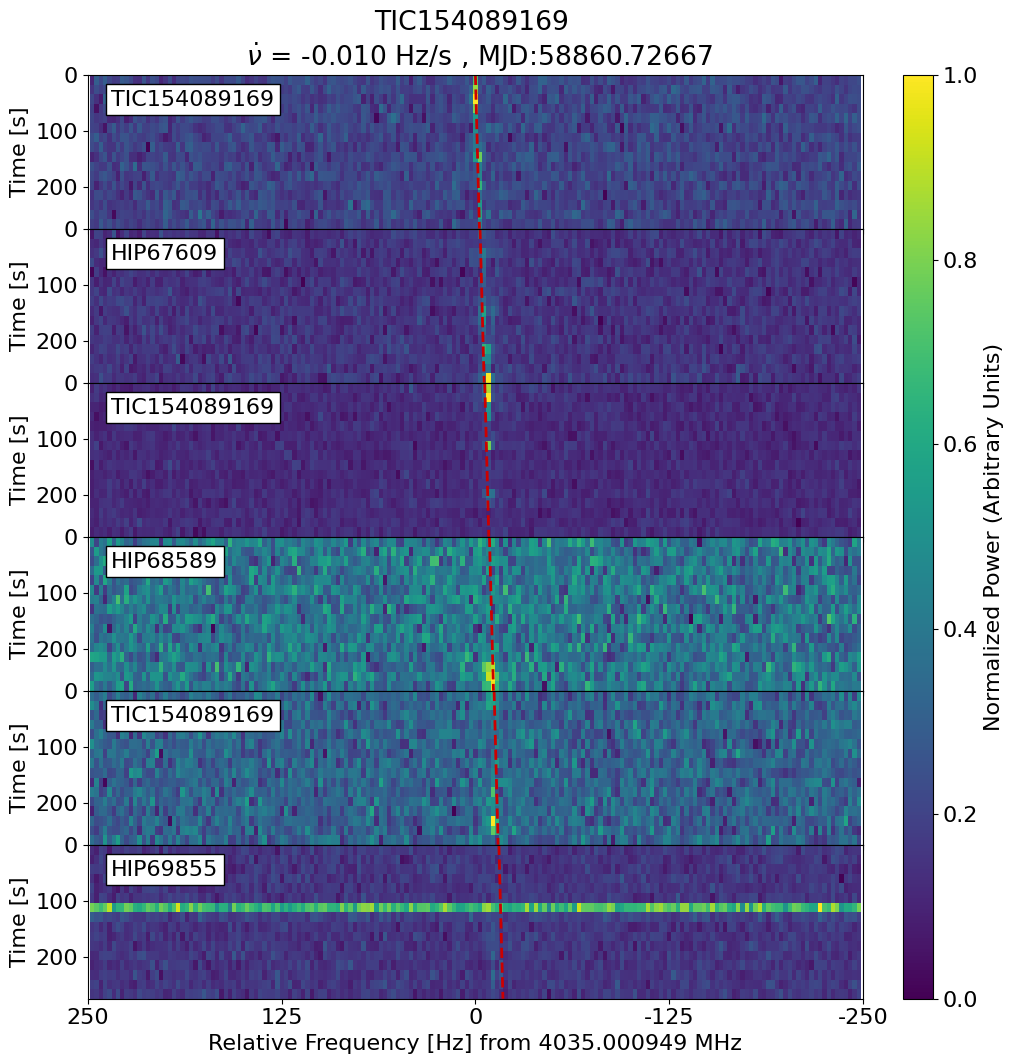}
  \label{fig:best-candidate-CBand}%
}\qquad
\subfloat[X-Band]{%
  \includegraphics[width=.42\columnwidth]{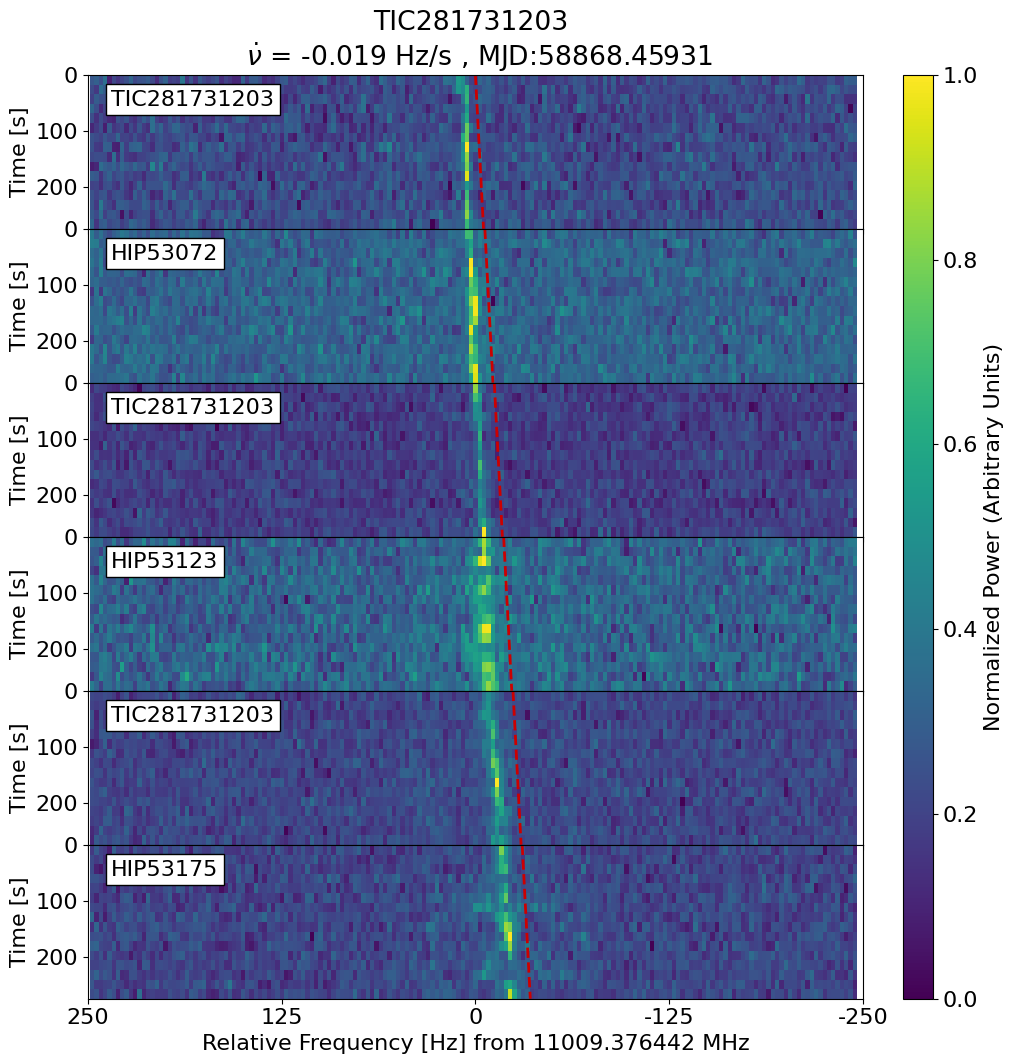}%
  \label{fig:best-candidate-XBand}%
}\qquad
\caption{\label{fig:best-candidates}Spectra of events found by \tseti\ found at each band.  The panels are ordered sequentially following the `ON'--`OFF' observation cadence.  The red dashed line indicates the drift rate computed by \tseti\ for the first hit in the candidate event (although the underlying algorithm searches a range of drift for matches in the other five observations). Note that in cases such as 
\reffig{fig:best-candidate-LBand} and \reffig{fig:best-candidate-SBand}, the red line, assuming constant Doppler drift, is a poor match for the changing drift rate of the detected signal. \label{fig:best-candidates}}
\end{figure*}

\section{Event Groups}\label{sec:appendix-event-groups}
\begin{figure*}[h]
\includegraphics[width=\linewidth]{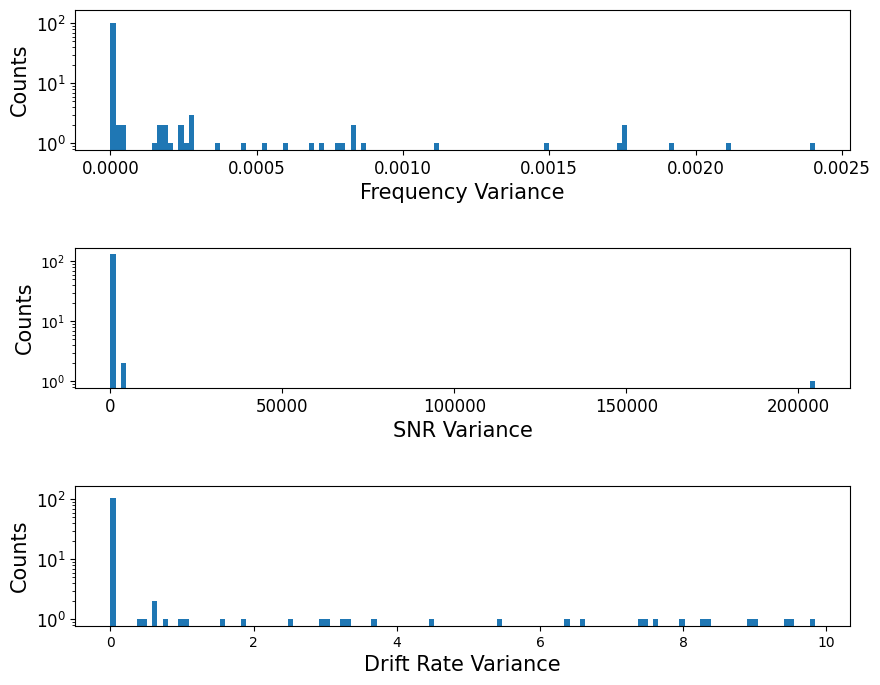}
\caption{Histogram showing the variance of frequency, \snr, and drift rate of all event groups at L-band.  All single source events (events with a standard distribution $\displaystyle \sigma < 1$) for frequency, \snr, and drift rate, were found to exist only at L-band.  We show the variance instead of the standard deviation since it is more effective in detecting outliers. \label{fig:lband-event-groups-variance}
}
\end{figure*}

\begin{figure*}[!h]
\centering
\begin{tabular}{cc}
\subfloat{
\includegraphics[width=0.5\columnwidth]{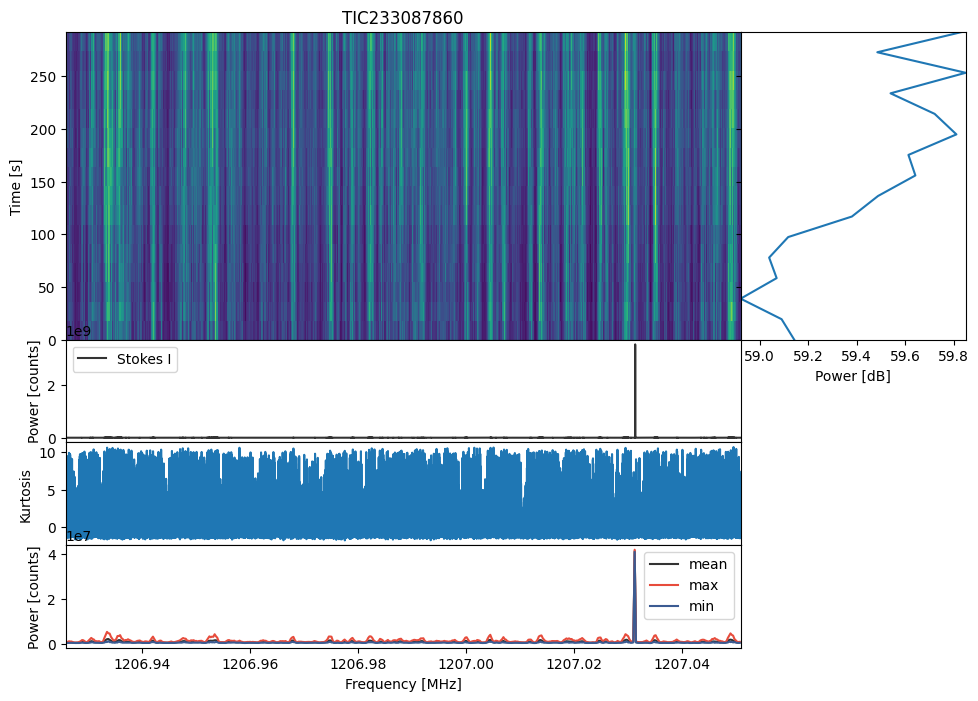}
\label{fig:single-source-1}
}\qquad
&
\subfloat{
\includegraphics[width=0.5\columnwidth]{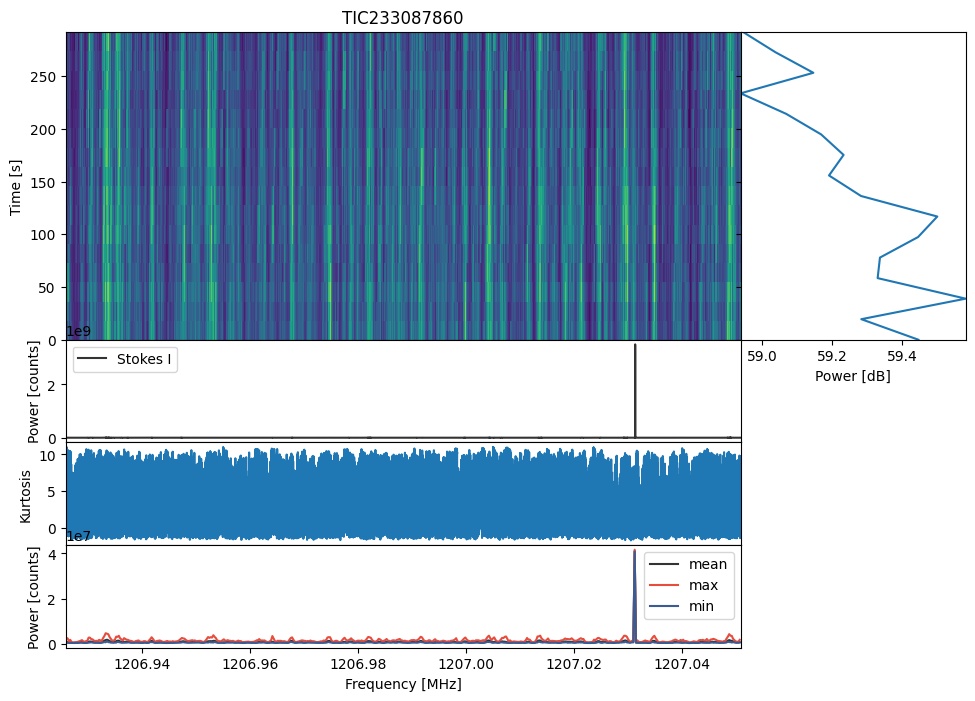}
\label{fig:single-source-2}
}\qquad
\\
\subfloat{
\includegraphics[width=0.5\columnwidth]{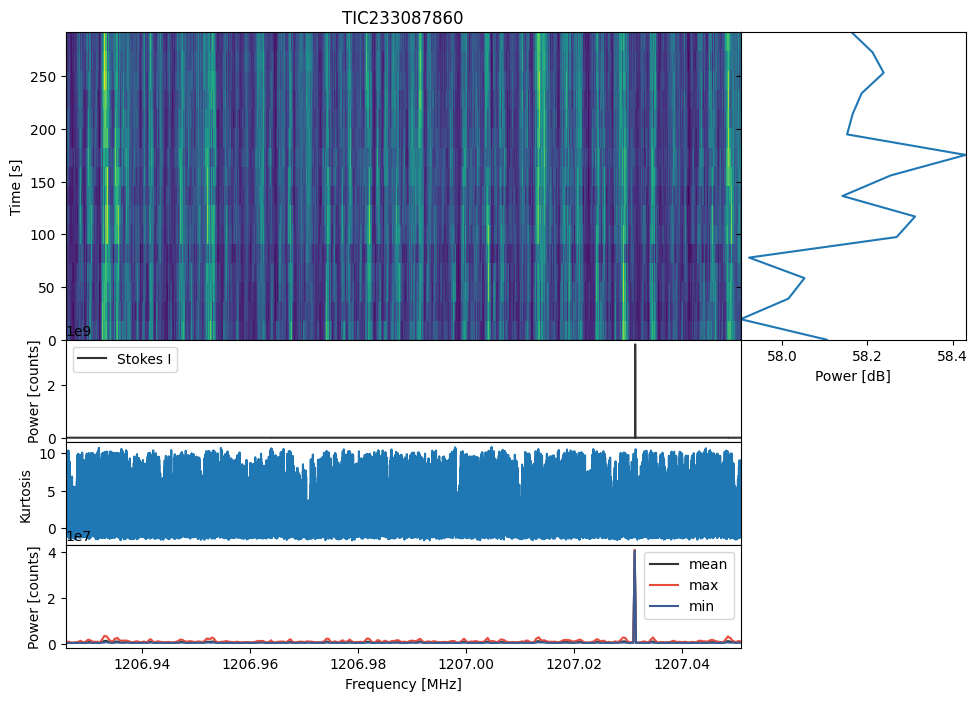}
\label{fig:single-source-3}
}\qquad
&
\subfloat{
\includegraphics[width=0.5\columnwidth]{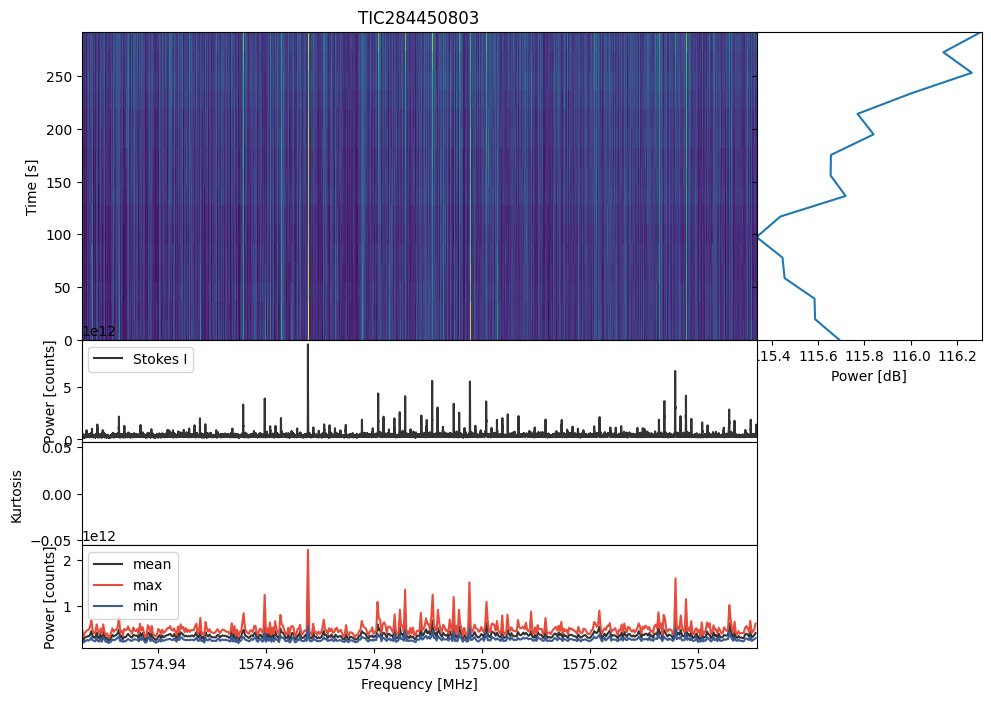}
\label{fig:single-source-4}
}\qquad
\\
\end{tabular}
\caption{\label{fig:single-sources-1} Examples of 125 kHz-wide, L-band, single-source event groups traceable to observations of TIC233087860 and TIC284450803.  Note how clustered signals exhibit nearly identical drift rates, \snr, and frequency.  Note the frequencies shown are both present in navigation satellite bands, and these observations appear to be badly affected by RFI.}
\end{figure*}

\begin{figure*}[!h]
\centering
\begin{tabular}{cc}
\subfloat{
\includegraphics[width=0.5\columnwidth]{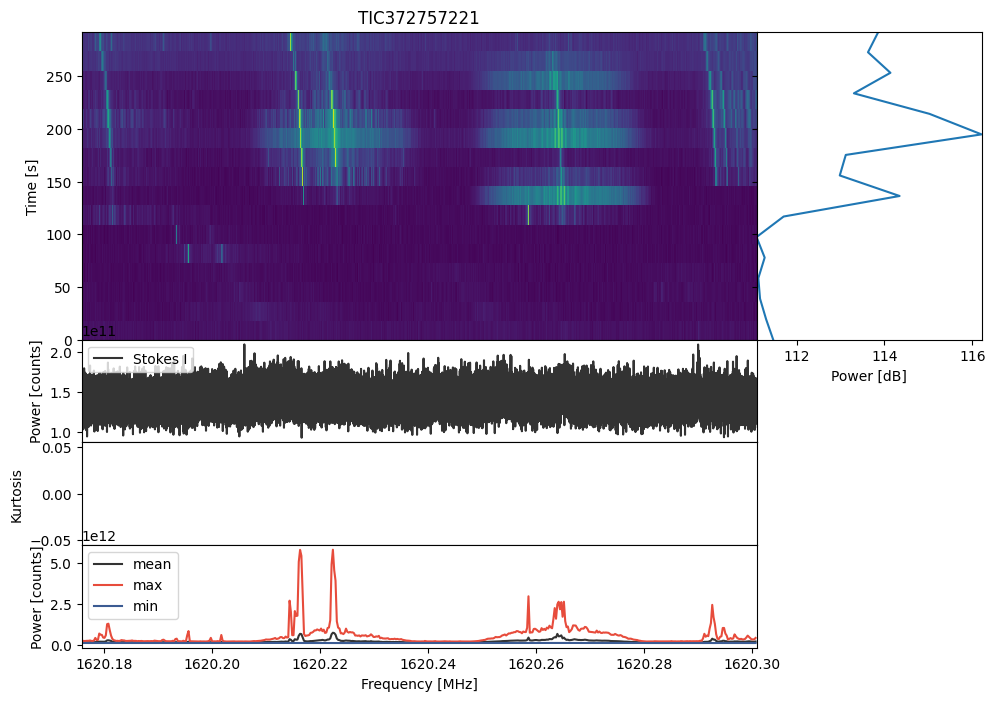}
\label{fig:single-source-5}
}\qquad
&
\subfloat{
\includegraphics[width=0.5\columnwidth]{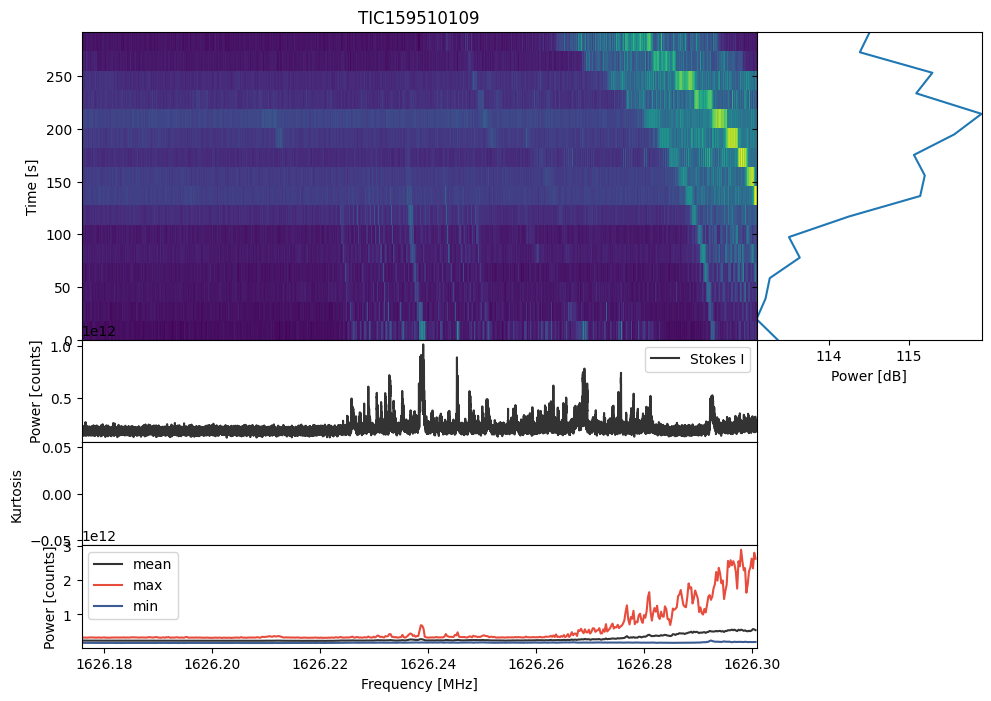}
\label{fig:single-source-6}
}\qquad
\\
\subfloat{
\includegraphics[width=0.5\columnwidth]{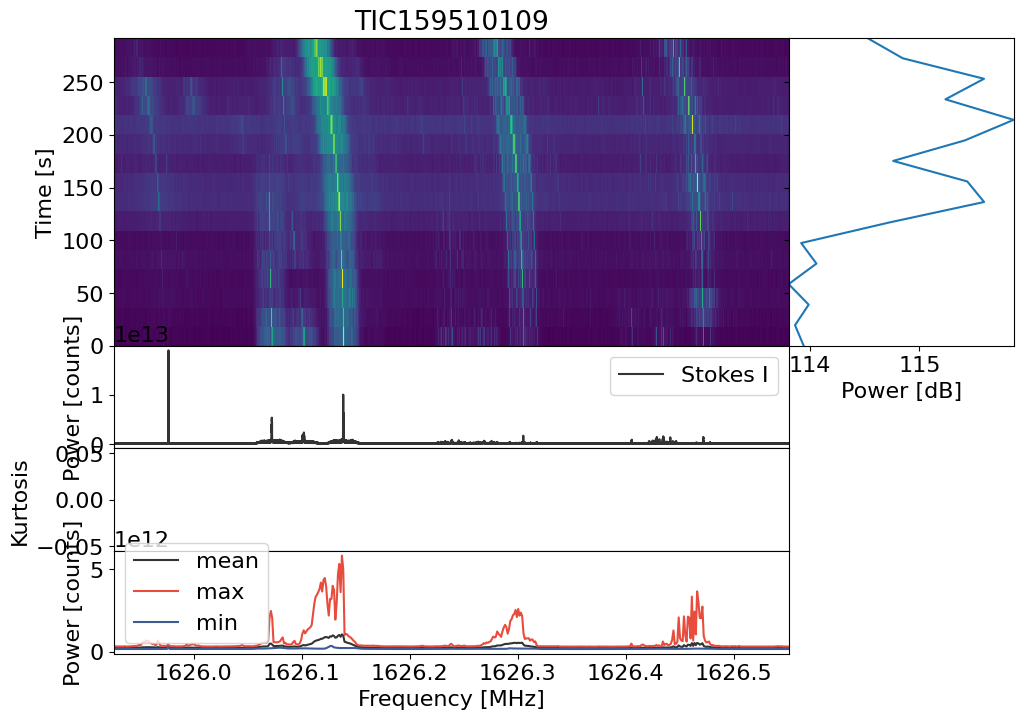}
\label{fig:single-source-7}
}\qquad
&
\subfloat{
\includegraphics[width=0.5\columnwidth]{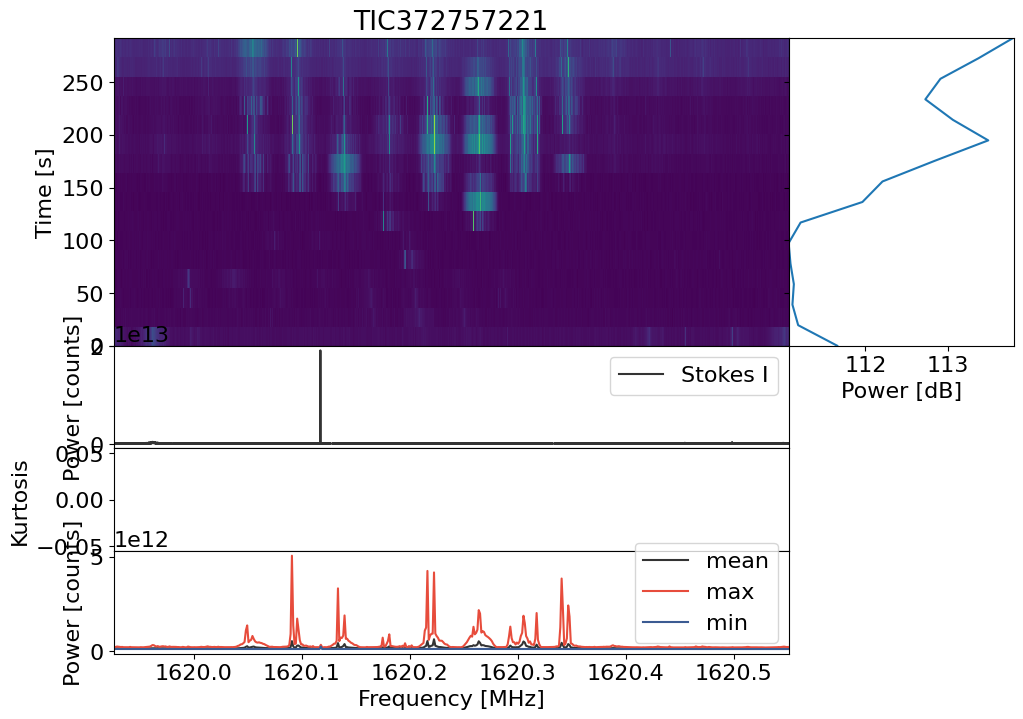}
\label{fig:single-source-8}
}\qquad
\\
\end{tabular}
\caption{\label{fig:single-sources-2} Examples of 125 kHz-wide, L-band, single-source event groups traceable to observations of TIC372757221 and TIC159510109.  Note all shown events have frequencies in the Iridium satellite band.}
\end{figure*}

Event groups are 125\,kHz-wide frequency ranges candidates which contain candidates identified by the \tseti\ \texttt{find\_event} method.  Since signals emanating from a single source are likely to be clustered in frequency and exhibit near-identical \snr\ and drift rates, we consider event groups possessing a standard deviation less than 1 in frequency, \snr, and drift rate as single sources.  The distribution of these variances is shown in \reffig{fig:lband-event-groups-variance}.  In this way, technosignatures are those signals originating from single sources unattributable to RFI.  There were no single-source signals which could not be attributed to RFI, but include sample plots of candidate technosignature sources shown in \reffig{fig:single-sources-1} and \reffig{fig:single-sources-2}.  All single-source event groups are traceable to 5 unique targets.

\end{document}